\documentclass{article}
\usepackage{amsmath,amssymb}
\usepackage[utf8]{inputenc}
\usepackage[english]{babel}  
\usepackage{hyperref}
\hypersetup{colorlinks=true, linkcolor=blue, citecolor = blue, urlcolor=blue}
\usepackage[ruled,vlined]{algorithm2e}    
\usepackage{url}
\usepackage{booktabs}       
\usepackage{amsfonts}      
\usepackage{nicefrac}       
\usepackage{microtype}
\usepackage[top=1 in,bottom=1in, left=1 in, right=1 in]{geometry}
\usepackage{graphicx}
\usepackage{subfig}
\usepackage{float}
\usepackage{wrapfig}
\usepackage{amsmath}
\usepackage{amssymb}
\usepackage[round]{natbib}
\usepackage{tabularx}
\usepackage[export]{adjustbox}
\usepackage{multicol}
\usepackage{color}
\usepackage{multirow}
\usepackage[thinlines]{easytable}

\title{Exploring the big data paradox for various estimands using vaccination data from the global COVID-19 Trends and Impact Survey (CTIS)}
   
\author{Youqi Yang$^1$, Walter Dempsey$^1$, Peisong Han$^1$, Yashwant Deshmukh$^2$, \\ Sylvia Richardson$^3$, Brian Tom$^3$, Bhramar Mukherjee$^{1 \ast}$} 

\date{%
    $^1$Department of Biostatistics, University of Michigan, Ann Arbor, MI, USA.\\%
    $^2$Center For Voting Opinions and Trends in Election Research, Noida, India.\\%
    $^3$MRC Biostatistics Unit, University of Cambridge, Cambridge, UK.\\
    $^\ast$Corresponding author. Email: bhramar@umich.edu. \\[2ex]%
}

\begin{document}

\maketitle

\begin{abstract}
    Selection bias poses a challenge to statistical inference validity in non-probability surveys. This study compared estimates of the first-dose COVID-19 vaccination rates among Indian adults in 2021 from a large non-probability survey, COVID-19 Trends and Impact Survey (CTIS), and a small probability survey, the Center for Voting Options and Trends in Election Research (CVoter), against benchmark data from the COVID Vaccine Intelligence Network (CoWIN). Notably, CTIS exhibits a larger estimation error (0.39) compared to CVoter (0.16). Additionally, we investigated the estimation accuracy of the CTIS when using a relative scale and found a significant increase in the effective sample size by altering the estimand from the overall vaccination rate. These results suggest that the big data paradox can manifest in countries beyond the US and it may not apply to every estimand of interest.
\end{abstract}

\section{Introduction}

The use of internet-based surveys has become increasingly common in population-based research. Survey researchers can administer questionnaires through internet-based platforms at a relatively low cost to a potentially large number of respondents \citep{Keiding2016}. However, these non-probability surveys may suffer from selection bias, which can lead to invalid inference, especially with a very large sample \citep{Bethlehem2010}. In a total survey error framework, one such source of the bias is coverage bias, as respondents to online questionnaires are limited to those with access to the Internet \citep{Matthias2009}. The distinction between the sampling frame and the target population can impact the external validity of any statistical inference. Non-response bias is another source of bias that arises when respondents and non-respondents differ significantly in the outcome of interest \citep{Groves2010}. Online surveys have reportedly low response rates. In the absence of a standard probabilistic design, respondents are self-selected, which can negatively impact representativeness \citep{Keiding2016}. For instance, those who are more concerned about SARS-CoV-2 infection may be more likely to respond to a COVID-19-related questionnaire \citep{Nguyen2022-ss}. With a large sample size resulting in very small variance, and biases that do not diminish with sample size, the squared bias term dominates the mean squared error (MSE) of the estimator. Thus, for a big self-selected sample as obtained from an internet-based survey, the focus of our statistical thinking should be controlling bias, rather than minimizing variance \citep{Meng2018}.
\\
\\
For estimating coronavirus vaccine uptake, the official population-level data collected by government agencies are often publicly available as a reliable benchmark \citep{CDC2023, GovernmentOfIndia, Dong2020-or}. The existence of such gold standard data gives researchers the opportunity to quantify estimation error of vaccine uptake estimates from other surveys. For example, using reports from the US Centers for Disease Control and Prevention (CDC) as the benchmark, researchers found the COVID-19 Trends and Impact Survey (CTIS), which is a big non-probability survey administered through Facebook, produced a more biased estimate of the first-dose vaccination rate among US adults when compared to a small probability survey \citep{Bradley2021-io}. The study by \cite{Bradley2021-io} used a decomposition framework proposed by \cite{Meng2018}, describing what is now known as the ``Big Data Paradox''. This framework enables researchers to decompose the estimation error into three components: (i) data defect correlation (ddc), (ii) data deficiency, and (iii) inherent problem difficulty. The ddc quantifies the extent of selection bias by measuring the correlation of the value of the response with the response indicator. This framework shows that even with vast amounts of data, a small ddc can induce a huge estimation error and dramatically reduce the effective sample size. Based on the study by \cite{Bradley2021-io}, one may naturally ask the question of whether studies like CTIS which were done through two years of the pandemic have any value at all. In this paper, we aim to illustrate cases where the CTIS survey may still be helpful in capturing relative changes in the outcome trajectory (namely vaccination data).
\\
\\
Large online surveys with individual-level data on COVID vaccination can provide valuable additional information beyond the overall vaccination rate including the variation of vaccination rates across socio-demographic factors, health-related behaviors, and features of a specific community that may not be captured by the aggregate data released by the official national statistics \citep{Salomon2021-xe, Kush2022, Lupton-Smith2022-lr}. They may also be able to capture changes over time as a result of policy interventions. The more granular level data from surveys can help identify vulnerable social sub-groups with less access to the vaccine and less trust in the healthcare system \citep{McCabe2021-dx, Nguyen2022-sd, McNaghten2022-au, Wang2021-kt}. This may lead to targeted intervention and campaign strategies for specific groups. Thus, despite the reality of the big data paradox for estimating the overall vaccination rates or more broadly overall prevalence, these non-probability surveys when used appropriately may help us reveal relative/comparative patterns.
\\
\\
This paper has two main objectives. First, as a parallel to the study by \cite{Bradley2021-io} we aim to evaluate the accuracy of the CTIS in estimating the first-dose COVID-19 vaccination rate among the Indian adult population. Our goal is to investigate whether the big data paradox that has been explored in the US also holds for a populous country like India with a completely different fabric of society and healthcare system. One unique aspect of our study is that in addition to CTIS, we used a population-based sample survey conducted via computer-assisted telephone interviewing (CATI) in India during the contemporary timeframe. We compared both of these COVID-tracker surveys to the gold-standard official data released by the Government of India and displayed their similar features. Second, we sought to determine whether a survey that is substantially biased in estimating the population average of the target variable could still be useful if the estimand is changed to: (a) the successive difference or relative difference in the population average over time; (b) differences in population averages of two subgroups (such as gender); and (c) the population average of a correlated variable (such as vaccine hesitancy) where gold standard population level official statistics may not be available. Since the US CTIS produced a biased estimate of the population-level first-dose COVID-19 vaccination rate \citep{Bradley2021-io}, we studied the above three problems using the US CTIS data. We also investigated the estimand (a) using the Indian CTIS data, while other issues in India were not explored due to limited data availability.
\\
\\
The rest of the paper is organized as follows. In Section 2, we presented our actual data analysis results, which indicate that the big data paradox also applies to India, as previously noted in the study by \cite{Bradley2021-io} for the US. We also demonstrate that the estimation error and reduction in effective sample sizes for these estimands compared to the benchmark data are much less than our previous findings in India and the results presented in the study by \cite{Bradley2021-io} in the US for estimating the population vaccination rates. In Section 4.1, we introduced the data sources we used for both India and the US. In the remaining part of Section 4, we focus on illustrative examples with the new estimands (a)-(c) where we believe the CTIS survey could be helpful. We first reviewed the framework of error decomposition proposed by \cite{Meng2018} and adapted to the relative estimands described in (a) and (b). In addition, we described the methods we applied to the estimands mentioned in (c). 

\section{Results}

\subsection{Characteristics of the respondents}

\paragraph{India} Table 1 summarizes the unweighted and weighted demographics from both the CTIS and the CVoter survey in India, compared to the latest Census reports. The weights used in both surveys, as described in Section 4.1, aimed to address the differences between the samples and the general adult population. However, the CTIS weights only partially corrected the bias regarding age and gender, resulting in the weighted samples still deviating from the general population. In particular, the proportion of individuals aged between 25 and 44 years old or men in the weighted samples of the CTIS (53\% and 66\%, respectively) was closer to that in the general adult population (47\% and 52\%, respectively) when compared to the unweighted samples (63\% and 84\%, respectively).
\\
\\
In terms of education and rurality, which were considered in the CVoter weights but not in the CTIS weights, the weighted proportions in the CVoter survey were more in line with the census data. Specifically, the weighted proportion of individuals with at least a college education or residing in an urban area in the CVoter survey (13\% and 30\%, respectively) was closer to the corresponding figures in the census data (13\% and 31\%, respectively), whereas the CTIS survey showed substantially higher proportions (74\% and 75\%, respectively).

\paragraph{India CTIS vs. US CTIS} Table 2 highlights a comparison of the demographic characteristics of the sampling frame (adult Facebook Active User Base) in India and the US with those of the weighted respondents from the CTIS \citep{facebook2021}. We used the latest Census data for India and the March 2019 CPS Supplement for the US as demographic benchmarks. The segment of adults that were Facebook active users in 2021 was smaller in India than in the US (40\% vs. 87\%). Among those who used Facebook actively, an overwhelming majority were men (76\%) in India, whereas, in the US, the majority were women (56\%). 
\\
\\
The weighted CTIS respondents in both India and the US differed from the demographic benchmarks regarding the distribution of certain variables. However, the deviation was more pronounced in India than in the US. Specifically, the discrepancy in the proportion of female participants between the weighted respondents and the general adult population in India was larger than that in the US (14\% vs. 1\%). 

\subsection{Big Data Paradox}

\subsubsection{Vaccine uptake in India}

Both the CTIS and the CVoter survey reported a higher percentage of the adult population in India receiving their first dose of the COVID-19 vaccine from May 16 to September 18, 2021, compared to the CoWIN benchmark (Figure 1A). The CTIS had a larger sample size (median 25,000 vs. 2,700) than the CVoter but its estimate of the vaccination rate was less accurate than the estimate generated by the CVoter survey. The range of estimation errors from the CTIS was between 25 to 45 percent, whereas from the CVoter survey, the range was between 5 to 20 percent (Figure 1B). This finding is consistent with the Big Data Paradox, which suggests that a larger non-probability survey (CTIS) may produce more biased estimates with narrower confidence intervals than a smaller designed survey (CVoter). 
\\
\\
Considering the three components of the error (namely inherent problem difficulty, data deficiency, and ddc) in the formula proposed by \cite{Meng2018}, the values for inherent problem difficulty in both surveys were identical (Figure 1C) since we assumed the CoWIN data as the reference standard in both cases. However, the data deficiency values in the CVoter survey were comparatively higher than those in the CTIS (median 588 vs. 224; Figure 1D), which can be attributed to differences in the original sample sizes (median 2,700 vs. 25,000). The ddc magnitudes in the CVoter survey were lower than those in the CTIS (median 0.0005 vs. 0.0040; Figure 1E). This may be due to the fact that the weighted respondents in the CVoter survey were more representative of the general population and the recording mechanism may not be related to the vaccination status.
\\
\\
The median weekly effective sample size of the CTIS and the CVoter survey was found to be 2 and 9, respectively (Figure 1F). One would not anticipate that an SRS survey with a sample size of 2 could have the same MSE as observed from a survey with an original sample size of 25,000 (the CTIS). The effective sample size in the CTIS was reduced by approximately 99.99\% compared to the original sample size (Figure 1G). The dramatic reduction of the sample size offers compelling evidence for the Big Data Paradox, which states that a non-probability survey with a large sample size can be misleading when using the original estimand.

\subsubsection{Comparing vaccine uptake between India and the US}

We evaluated the vaccine uptake in the US from February 7 to May 15, 2021, against those in India between May 16 and September 18, 2021, regarding the estimation error and its three components. It is important to note that the time frame of our investigation in the US differs from that of the study by \cite{Bradley2021-io}, as described in Section 4.2. The Indian and US CTIS both overestimated the vaccine coverage among adults when compared to the benchmark data in each country (Figure S1). Nevertheless, the CTIS in India had larger estimation errors than the CTIS in the US over time (median 0.39 vs. 0.16; Table 3). The larger difference between the estimated and benchmark values in India may be caused by the higher values of data deficiency in India compared to the US (median 224 vs. median 34). Although India has a larger general population, the original sample sizes were significantly smaller than the US. The inherent problem difficulty levels in the Indian CTIS were comparable to those of the US CTIS, as a result of our decision regarding study duration (median 0.47 vs. 0.49). The values of ddc were smaller in the Indian version than in the US version (median 0.0040 vs. 0.0079). This indicates that the selection mechanism of the respondents from the US was more biased. Previously, we found that the sampling frame in the US version covered a larger proportion of the entire population, and the respondents in the US version were more representative of the general adult population considering certain variables. However, neither of these aspects can play a dominant role in the selection mechanism, which is in agreement with prior research \citep{Bradley2021-io}.
\\
\\
Regarding small probability surveys, it is noteworthy that the vaccination rate estimates obtained from the CVoter survey did not align as well with the population averages reported in the benchmark, in contrast to the Axios-Ipsos survey (Table 3). Despite having similar data deficiency and inherent problem difficulty, the CVoter survey had much larger estimation errors compared to the Axios-Ipsos survey (median 0.16 vs. 0.01). This was owing to the fact that the CVoter survey had significantly larger values of ddc (median 0.0005 vs. 0.0001). Although the CVoter survey employed a probability-based sampling and had respondents that seemed representative, it still retained some selection biases.

\subsubsection{Vaccine uptake in other countries}

In addition to India and the US, we expanded our analysis of the estimation error of vaccine uptake from the CTIS to 85 other countries. The benchmark data were sourced from the Johns Hopkins Coronavirus Resource Center \citep{Dong2020-or}. We developed an R shiny app to showcase the results (\url{https://3ogdqc-youqi-yang.shinyapps.io/LLPinVaccine/}). Developed countries in Europe showed a lower estimation error, lower ddc, and a higher effective sample size over time in comparison with low- and middle-income countries, especially those in Africa. This suggests that the selection process for the respondents in developed countries was less influenced by bias. Across various countries, the disparity between the CTIS estimates and the benchmark was widely noted, implying that the observations on the Big Data Paradox are not exclusive to particular locations. 

\subsection{Exceptions to the Big Data Paradox}

\subsubsection{Successive difference and relative successive difference in vaccine uptake}

During the time frame of February 7 to May 15, 2021, we checked the CTIS estimates of the vaccination rate difference and relative difference at consecutive time points, using the method outlined in Section 4.4. Compared to the CDC benchmark, the median estimation error of the difference and relative difference from the US CTIS was 0.007 and -0.010, respectively. The estimation error for the vaccination rate per week was 0.128 (Figure 2A). The estimands based on the successive difference and relative successive difference yielded considerably larger effective sample sizes compared to the original estimand (median 10,113; 18,012; 15; Figure 2B), representing a considerably higher proportion of the original sample size (median 3.74\%; 7.59\%; 0.01\%). 
\\
\\
In a similar manner, we examined the successive difference and relative successive difference in vaccination rates as estimated by the Indian CTIS from May 16 to September 18, 2021. In comparison with the CoWIN benchmark, we observed a median estimation error of -0.003 for the difference and -0.039 for the relative difference. The estimation error for the weekly vaccination rate was 0.390 (Figure 2C). Notably, the estimands based on successive difference and relative successive difference showcased substantially larger effective sample sizes when compared to the original estimand (median 1,922; 2,929; 2; Figure 2D), accounting for a significantly higher proportion of the original sample size (median 10.18\%; 14.94\%; 0.01\%). The findings in the US and India indicate that the CTIS can provide a more reliable estimate of the absolute and relative number of newly vaccinated people, compared to its estimate of cumulative vaccinated people.

\subsubsection{Gender difference in vaccine uptake}

Using the CTIS and the CDC benchmark data from February 7 to May 15, 2021, we investigated the estimated difference in adult vaccine uptake between males and females in the US, applying the methodology explained in Section 4.5. While both the CTIS and the benchmark data indicated that women were more likely to be vaccinated during the study period, the CTIS underestimated the magnitude of the gender gap in vaccination rates compared to the benchmark (Figure S2). 
\\
\\
Compared with the overall vaccination rates, the gender differences exhibited relatively smaller estimation errors (median -0.03 vs. 0.18; Figure 3A). Of the three components of the estimation error, the inherent problem difficulty was higher for the gender difference than for the overall vaccination rate (median 1.14 vs. 0.48; Figure 3B). The values of data deficiency were identical in both scenarios (Figure 3C). In terms of ddc, the gender difference estimates had smaller magnitudes (median -0.006 vs. 0.103), resulting in significantly greater effective sample sizes (median 1,999 vs. 7) and less loss of the original sample size (median 98.91\% vs. 99.99\%), compared to the overall vaccination rate estimates (Figure 3D, Figure 3E, and Figure 3F). The results suggest that a survey that is biased in estimating the overall rate can still provide a more accurate estimate of the difference among subgroups.

\subsubsection{Using vaccine uptake as auxiliary information for vaccine hesitancy}

We analyzed vaccine hesitancy among US adults from January 10 to May 15, 2021, with the aid of vaccine uptake as an auxiliary variable, utilizing the approach detailed in Section 4.6. The Axios-Ipsos survey showed a strong correlation between vaccine hesitancy and vaccine uptake ($\rm pseudo \: R^2 = 0.86$). The CTIS survey underestimated the extent of vaccine unwillingness among US adults when model-assisted estimates from the Axios-Ipsos survey were used as benchmarks (Figure 5A). Vaccine hesitancy decreased over time in both surveys as the vaccine program progressed (Figure 5A), which aligns with findings from previously published research \citep{Daly2021-xe}. The estimates using vaccine uptake as auxiliary information exhibited smaller estimation errors than the original estimates from the Axios-Ipsos survey (median -0.01 vs. -0.13; Figure 5B). The effective sample sizes were larger in the model-assisted CTIS estimates than those in the original CTIS estimates (median 7896 vs. 13; Figure 5F). When an auxiliary variable that is highly correlated with the variable of interest is available, it can help to improve estimation accuracy. 

\section{Discussion}

Our study serves as both a parallel and an extension to previous research on COVID-19 vaccination rates among adults in the United States using CTIS data \citep{Bradley2021-io}. In Section 4.2, akin to the work by \cite{Bradley2021-io}, we demonstrate that a large non-probability survey (namely CTIS) produces more biased estimates of vaccination uptake among Indian adults in comparison to a smaller probability survey (namely CVoter). These findings emphasize that the big data paradox holds for data from low- and middle-income countries (LMICs), where the weighted proportion of the respondents with at least a college education or residing in an urban area was notably higher compared to the census data. In comparison to the US CTIS results during a similar period of benchmark vaccination rates, the Indian CTIS exhibited a higher value of estimation error over time. The Indian version of the CTIS suffered more from data deficiency compared to the US version, while it had a smaller value of ddc. Additionally, our comparison of small probability surveys in India (namely CVoter) and the US (namely Axios-Ipsos) shows that the former did not estimate the overall vaccination rates that closely matched the benchmark, unlike the latter.
\\
\\
In contrast, in Section 4.3, we intend to understand if the CTIS data is useful for estimating relative trends as opposed to the overall vaccination rates. Using the same data sources as in the study by \cite{Bradley2021-io}, we observe that despite the obvious selection bias, the CTIS can provide more accurate estimates of (a) the successive difference and relative successive difference in vaccination rates, (b) gender differences in vaccination rates, and (c) vaccine hesitancy when using vaccination uptake as an auxiliary variable. For instance, the effective sample sizes for the successive difference were approximately 700 times greater than the effective sample sizes for the overall vaccination rate. These results offer a more optimistic perspective of the CTIS data and similar large non-probability samples, indicating that the Big Data Paradox is not inevitable for every estimand one may be interested in. Instead, we suggest that researchers study each estimand on its own for analysis while analyzing self-selected samples. For instance, considering the lack of gender-specific vaccination data provided by the benchmark in India, the CTIS estimates can offer valuable insights into the disparity in vaccination rates between different genders (Figure S3). \cite{kundu2023framework} compared different weighting approaches to obtain bias-reduced inference when dealing with data that have substantial selection bias. Here, we did not pursue a more in-depth weighted analysis as our objective was primarily to illustrate the prevalence of the big data paradox across different estimands. It is important to acknowledge that designing stratified weights for vaccine uptake is challenging due to the presence of numerous unmeasured individual variables that can potentially influence the response.
\\
\\
In order to assess the effectiveness of the relative scale estimation, it is beneficial to explore alternative methods beyond direct estimation error calculations. Prior studies have involved the utilization of randomized surveys to enhance the statistical precision of local estimates pertaining to COVID-19 infections \citep{Nicholson2022-gj}. Here, we performed a method of change point detection as a reference for future studies. We intended to detect the time when a significant change occurs in the vaccination rate estimates obtained from the benchmark data, by using the survey data. These shifts could indicate factors such as a surge in infections or the implementation of new interventions. Both the benchmark and the CTIS have multiple observations over the study period, so we considered both sets of data to be time series data. To address this question, we applied an established change point detection method for time series data, specifically utilizing the Bayesian change point analysis method from the bcp package in R \citep{Wang2015}. For each time point, we can calculate the posterior probability of it being a change point. We considered a time period to be a potential interval for containing a change point if all time points within it have a posterior probability greater than 0.6. Our detection of the change point in the new vaccination rate among the US adult population from the CTIS and the CDC benchmark covered the period between January 18 and November 28, 2021. The time series data from the CTIS and the CDC identified four periods with potential change points that overlapped (Figure S4). The time periods detected by the CTIS were from February 26 to March 2, from April 9 to April 14, from April 23 to April 27, and from May 22 to May 24, 2021. Similarly, the time periods detected by the CDC benchmark were from February 19 to March 2, from April 9 to April 14, from April 23 to April 27, and from May 21 to May 25, 2021. Notably, there was also consistency in the direction of changes observed by both sources. Specifically, both sources observed a rapid surge in new vaccination rates among US adults between late February and early April 2021, followed by stepwise decreases in the three subsequent change point periods (Figure S4). This surge aligned with the Delta-variant wave in the United States \citep{Diesel2021-ls}, and may have been influenced by the increase in the availability of the COVID-19 vaccine in late February \citep{fda-eua-20210227}, as well as increasing vaccine eligibility \citep{ourworldindata-2023}. Despite differences in overall estimates between the survey and the benchmark, the survey data proved valuable in detecting the timing of change points. This ability to identify critical shifts in vaccination rates is essential for policy evaluation and decision-making. 
\\
\\
This study has several limitations. First, our comparison of estimator errors in probabilistic and non-probability surveys was based solely on the framework of error decomposition proposed by \cite{Meng2018}. Other methods have been suggested to quantify non-ignorable selection mechanisms in non-probability surveys, including using indexes \citep{Andridge2019-bq, Little2020-zf, Boonstra2021-ji}. Future research could investigate the selection bias by utilizing these alternative methods. Second, as highlighted in Section 4.1.3, potential measurement errors were present in both the CDC and CoWIN benchmarks when recording vaccination doses. Although we conducted a corresponding sensitivity analysis, a small deviation in the assumed benchmark for vaccination rate could result in a change in estimation error, along with the contributing factors of the estimation error. Third, in the framework of error decomposition from the study by \cite{Meng2018}, we assumed that the response in the survey represented the true vaccination status of the respondent. However, it is possible that an unvaccinated individual might report being vaccinated due to social pressures, leading to differential misclassification. Therefore, it is crucial to account for uncertainty when utilizing this method, thus presenting a potential direction for future investigation. \cite{Dempsey2020} has proposed approaches for adjusting these measurement errors when applying error decomposition. Finally, the error decomposition proposed by \cite{Meng2018} is restrictive since it can only be applied to means. Further research can explore similar ideas in other parameters such as sample quantiles. 
\\
\\
However, for countries like India, where limited individual data are available at the national level, one should do the best possible analysis of the data source like the CTIS. We hope our work will be helpful to enhance the practitioners' understanding of the big data paradox and the alternatives that are available. We also hope our R shiny app and accompanying codes will encourage users to explore the global version of the CTIS as most of the published analyses of the CTIS are US-centric.

\section{Materials and Methods}

\subsection{Data source}

\subsubsection{Non-probability survey}

\paragraph{COVID-19 Trends and Impact Survey (India and the US)} Meta (formerly Facebook) conducted the COVID-19 Trends and Impact Survey (CTIS) through their social media platform from April 2020 to June 2022 in partnership with the University of Maryland (UMD) and Carnegie Mellon University (CMU) for global and US versions, respectively \citep{Astley2021-aq, Salomon2021-xe}. A stratified random sample of adults from the Facebook Active User Base (FAUB) was invited daily to complete a cross-sectional questionnaire on COVID-related symptoms, exposures, outcomes, mental health, economic security, and demographics. The continuously adapted survey waves first launched queries about coronavirus vaccine uptake in January 2021. A potentially attractive feature of the CTIS is its large sample size. India and the US survey data yielded approximately 25,000 and 250,000 respondents per week respectively, in 2021. Facebook designed weights to adjust for the differences between the respondents and the general population in two components: non-response weights and post-stratification weights \citep{Barkay2020}. For non-response weights, Facebook utilized an inverse propensity score weighting method to address the potential correlation between missingness and auxiliary information obtained from the user profile. This makes the survey respondent population more representative of the FAUB. For post-stratification weights, Facebook adjusted weights to match the respective national census benchmark in terms of age, gender, and region. This adjusted for those who are not actively on Facebook or do not have internet access and made the survey more representative of the source population per country. Aggregated weighted responses are publicly available \citep{Fan2020, Reinhart2021-yh}. The US microdata was downloaded from CMU's repository for daily tables with estimates and weights \citep{Reinhart2021-yh}. In our study, we primarily used aggregated weighted data. However, for the new estimand (b), we incorporated individual-level responses with weights provided by Facebook.

\subsubsection{Probability Survey}

\paragraph{CVoter COVID-19 Tracker Survey (India)} The Center for Voting Options and Trends in Election Research (CVoter) administered the COVID-19 Tracker survey to measure symptoms and attitudes toward COVID-19 among Indian adults \citep{CVotermethod} starting in March 2020. The data available for analysis was last updated in July 2022. Team CVoter recruits a probability-based random sample of participants from the general public for a telephone interview per wave utilizing CATI. The survey introduced a vaccine uptake question in May 2021. The weekly average sample size was roughly 2,700 in the timeframe of our study. Analysis weights have been developed by Team CVoter to account for the differences between the respondents and the overall population in terms of demographics. Detailed reports were made to be publicly available on their website \citep{CVoter} whereas our study used the individual level data via collaboration with the CVoter team. Table 4 compares the study design characteristics of the two surveys we used for India. 

\paragraph{Axios-Ipsos Coronavirus tracker (US)} The analogously designed survey in the US as used in the study by \cite{Bradley2021-io} was the Axios-Ipsos Coronavirus tracker. Axios and Ipsos rolled out the bi-weekly survey to investigate COVID-19-like symptoms in the US through the online KnowledgePanel starting in March 2020. The data available for analysis were last updated in December 2022 \citep{Jackson2021}. The selection mechanism for the survey involves an address-based probabilistic sampling methodology employing the delivery sequence file from the United States Poster Service. Ipsos provides a tablet and Internet connection for the potential adult respondents who lack online connectivity, ensuring the sampling frame covers the entire adult population. The survey first inquired about vaccine uptake in December 2020. The number of respondents was roughly 1,000 per wave (bi-weekly) throughout the study. Ipsos has developed weights to account for the differences between the respondents and the US adult population in terms of demographics. The weighted estimates were sourced from the topline PDF documents published on the Ipsos website \citep{Jackson2021}. Table S1 compares design characteristics between the surveys we used for the US. 

\subsubsection{Benchmark Official/Government Data}

\paragraph{CoWIN (India)}  We obtained the administrative vaccination counts from the COVID Vaccine Intelligence Network (CoWIN) portal extracted via the \href{https://data.covid19bharat.org}{website}. CoWIN is an online portal for individual vaccine registration, playing a vital role in real-time nationwide vaccination monitoring \citep{GovernmentOfIndia}. In order to align with the survey population, we must restrict the CoWIN vaccination counts to those given to adults. Given that India extended vaccine eligibility to citizens under 18 years old only in 2022 \citep{TimesOfIndia2021} it is reasonable to utilize cumulative counts of the first-dose vaccine administration for the total population in 2021 as the gold standard counts. Potential recording errors could arise in the CoWIN database during the large-scale vaccination drive, such as merging dose certificates, unidentified records, and correction of vaccination dates. To account for potential imprecision in the benchmark data, we conducted a sensitivity analysis by multiplying the reported values from CoWIN by 0.9, 0.95, 1, 1.05, and 1.1. 

\paragraph{CDC (US)} The CDC benchmark data were extracted from reports disaggregated by age group and sex on the website \citep{CDC2023}. We used the total number of first-dose vaccines assigned to people aged 18 or older to match the survey population. We excluded those who did not report their age ($<1\%$) in the analysis. In our comparison of vaccine coverage between genders, we additionally excluded individuals who did not report their gender information ($<1\%$). The CDC has acknowledged a lag time between when an individual receives a vaccine and when it is reported to the CDC. We performed a similar adjustment for uncertainty as the CoWIN by scaling the CDC-reported numbers by a factor of 0.9, 0.95, 1, 1.05, and 1.1.

\subsubsection{Demographic data (India and the US)}

Population-based demographic data used to calculate weights in the CTIS and the CVoter survey in India were extracted from the United Nations (UN) Population Division 2019 World Population Projection and the latest Census, respectively \citep{UN2019, IndianCensus2011}. The standard demographic distributions employed for the CTIS and the Axios-Ipsos survey weighting in the US were downloaded from the March Current Population Survey (CPS) Supplement, in 2017 and 2019, separately \citep{CPS2017, CPS2019}. Adult population sizes were extracted from the same sources.

\subsection{Study period}

Different illustrative examples in this paper have different time periods of analysis. Our analysis for objective 1, understanding the estimation error decomposition for national vaccination coverage in India, considers survey data from May 16 to September 18, 2021. We chose May 16, 2021, as our starting point since the CVoter survey launched the vaccination uptake question for the first time on that date. We chose September 18, 2021, as our ending point since the CoWIN benchmark estimated the proportion of the adult population that had received at least one dose of vaccine as 60\% on that date. As a comparison, we studied the overall vaccination rate among US adults between February 7 and May 15, 2021. We consider these two periods comparable as the benchmark vaccination rates we observed at the beginning and end of our analysis period in the US were similar to those we observed in India (15\% and 60\%, respectively). For our second objective in studying new estimands, we analyzed the successive differences, relative successive differences, and gender-specific subgroup differences in vaccination rates in each country over the same period as our first objective. Finally, we conducted an evaluation of COVID-19 vaccine hesitancy for scenario (c) in the US, focusing on the period from January 10 to May 15, 2021, the initial phase of the vaccination rollout where hesitancy estimates were most relevant.

\subsection{Error decomposition of the sample mean}

\cite{Meng2018} provided a decomposition of estimation error in population mean and derived a formula for the resulting bias-adjusted effective sample size, $n_{eff}$. For the sake of completeness, we review the main ideas here. Let $n$ denote the size of the (potentially non-probability) sample coming from a finite population of size $N$. Let $\overline{Y}_n$ be the sample average of the variable $Y$, and let $\overline{Y}_N$ be the population average. The discrepancy between $\overline{Y}_n$ and $\overline{Y}_N$ can be decomposed into three components using the following mathematical equation:

\begin{align}
    \underbrace{\overline{Y}_n - \overline{Y}_N}_\textrm{Estimation error} = \underbrace{\rho_{Y, R}}_\textrm{Data defect correlation} \times \underbrace{\sqrt{\frac{N - n}{n}}}_\textrm{Data deficiency} \times \underbrace{\sigma_{Y}}_\textrm{Inherent problem difficulty}.
\end{align}

\noindent The first component ${\rho}_{Y, R}$ represents the population correlation between the variable $Y$ and a binary recording indicator $R$, where we let $R$ take the value $1$ if we have recorded a value of $Y$ in the sample and 0 otherwise. This measure is also called data defect correlation because it directly addresses the bias due to the selection/recording mechanism by quantifying whether being included in the analytic sample depends on the outcome $Y$. The second component $\sqrt{(N-n) \slash n}$ represents data deficiency. The smaller the proportion of the entire population recorded, the larger the $\sqrt{(N-n) \slash n}$, and the more challenging it becomes to estimate the $\overline{Y}_N$ accurately. The third component $\sigma_{Y}$, the population standard deviation of $Y$ represents the inherent problem difficulty by capturing random variation in $Y$. 
\\
\\
The advantage of this representation as opposed to a standard bias-variance decomposition of the mean-squared error (which is the squared version of the left-hand side in equation (1) by taking expectation with respect to the distribution of $R$) is that we can compare surveys with different recording mechanisms and sample sizes drawn from the same population for estimating the same target population quantity in terms of their accuracy.
\\
\\
In our investigation of the national coronavirus immunization rate, $n$ is the survey sample size, $N$ is the entire adult population size, and $\overline{Y}_n$ is the survey estimate of the vaccination rate. We have the rare and unique situation of knowing $\overline{Y}_N$ and $\sigma_{Y}$ from the national-level benchmark dataset. Thus for different survey mechanisms (probability or non-probability) the data defect correlation $\rho_{Y, R}$ can be estimated by plugging in the other values in equation (1). We denote the estimated value as $\hat{\rho}_{Y, R}$, which is subject to potential measurement errors in both the surveys and the benchmark data. This quantity is not estimable from just sample data where we only observe $R=1$. 
\\
\\
The effective sample size ($n_{eff}$) is defined as the size of a simple random sample (SRS) drawn from the same population that will produce the same MSE as observed in the survey of interest with selection mechanism $R$. The formulas for the observed MSE (${\rm MSE}_R$) and the MSE from the SRS (${\rm MSE}_{SRS}$) are as follows.
\begin{align*}
    {\rm MSE}_R = E_R[{\rho}_{Y, R}^2] \times \frac{N - n}{n} \times \sigma^2_{Y}.
\end{align*}
\begin{align*}
    {\rm MSE}_{SRS} = \Big[\frac{1}{N - 1}\Big] \times \frac{N - n_{eff}}{n_{eff}} \times \sigma^2_{Y}.
\end{align*}
\noindent By equating ${\rm MSE}_R$ and ${\rm MSE}_{SRS}$, we have:
\begin{align*}
    n_{eff} = \frac{\frac{n}{N-n} \times \frac{1}{E_R[{\rho}_{Y, R}^2]}}
    {(\frac{n}{N-n} \times \frac{1}{E_R[{\rho}_{Y, R}^2]} - 1) \times \frac{1}{N} + 1}.
\end{align*}

\noindent In our analysis of the vaccination rate, we approximate $E_R[{\rho}_{Y, R}^2]$ using the observed $\hat{\rho}_{Y, R}^2$. Since $N$ is large, we can assume that $1/N$ is close to 0, and simplify the formula for calculating the effective sample size as:

\begin{align}
    n_{eff} = \frac{n}{N - n} \times \frac{1}{\hat{\rho}_{Y, R}^2}.
\end{align}

\noindent Equation (2) shows that having a large sample size $n$ for the current survey does not necessarily guarantee accurate results if there is a substantial data defect correlation $\hat{\rho}_{Y, R}$. 

\subsection{Successive difference and relative successive difference in means}

While Section 4.3 compares the survey mean $\overline{Y}_n$ and the benchmark mean $\overline{Y}_N$, the method in this section compares survey estimates with the benchmark data when the targets are the difference and the relative difference in two means. Here we consider differences in vaccination rates over successive time periods as our target. This formulation is similar to the one in the study by \cite{Dempsey2020} for measuring the change in COVID-19 infection rates. Let $\overline{Y}_{n,t}$ be the survey average of the variable $Y$ at time $t$, and let $\overline{Y}_{N,t}$ be the population average at the same time $t$. In this scenario, the difference in the survey average of the variable $Y$ from time $t-1$ to $t$ is represented by $\overline{Y}_{n,t} - \overline{Y}_{n,t-1}$. The corresponding relative difference is represented by $(\overline{Y}_{n,t} - \overline{Y}_{n,t-1})\slash{\overline{Y}_{n,t-1}}$. The difference in the population average from time $t-1$ to $t$ is represented by $\overline{Y}_{N,t} - \overline{Y}_{N,t-1}$, and the corresponding relative difference is represented by $(\overline{Y}_{N,t} - \overline{Y}_{N,t-1})\slash{\overline{Y}_{N,t-1}}$. Let $\sigma_{Y_t}^2$ be the population variance of $Y$ at time $t$. Define the effective sample size $n_{eff}$ as the size of two SRS samples that can generate the observed MSE for the successive difference and relative successive difference. To simplify the analysis, we assume that both SRS samples have an equal sample size. We derive the following formulae for calculating the effective sample sizes in regard to the difference and relative difference at consecutive times. See Appendix 1.1 and 1.2 for a detailed calculation.

\begin{align}
    {\rm Difference}(n_{eff}) = \frac{\sigma_{Y_{t-1}}^2 + \sigma_{Y_t}^2}{[(\overline{Y}_{n,t} - \overline{Y}_{n,t-1})-(\overline{Y}_{N,t} - \overline{Y}_{N,t-1})]^2}.
\end{align}

\begin{align}
{\rm Relative \; Difference}(n_{eff}) = \frac{\Big(\frac{\overline{Y}_{N,t}}{\overline{Y}_{N,t-1}}\Big)^2\big[\frac{\sigma_{Y_{t-1}}^2}{(\overline{Y}_{N,t-1})^2} + \frac{\sigma_{Y_t}^2}{(\overline{Y}_{N,t})^2}\big]} {\big[\frac{\overline{Y}_{n,t} - \overline{Y}_{n,t-1}}{\overline{Y}_{n,t-1}} - \frac{\overline{Y}_{N,t} - \overline{Y}_{N,t-1}}{\overline{Y}_{N,t-1}}\big]^2}.
\end{align}

\noindent In order to represent the above two equations in  terms of the three components of the estimation error, we can employ equation (1) to decompose the estimation error at different time points, namely time $t-1$ and $t$. Denote $n_t$ as the sample size of the survey conducted during time $t$, and denote $N_t$ as the population size during time $t$. Additionally, denote ${\rho}_{Y_t, R_t}$ as the ddc at time $t$. Expanding the equations yields the following expressions. See Appendix 1.3 and 1.4 for a detailed calculation.

\begin{align*}
    {\rm Difference}(n_{eff}) = \frac{\sigma_{Y_{t-1}}^2 + \sigma_{Y_t}^2}{\big[\hat{\rho}_{Y_t, R_t} \times \sqrt{\frac{N_t - n_t}{n_t}} \times \sigma_{Y_t} -\hat{\rho}_{Y_{t-1}, R_{t-1}} \times \sqrt{\frac{N_{t-1} - n_{t-1}}{n_{t-1}}} \times \sigma_{Y_{t-1}}\big]^2}.
\end{align*}

\begin{align*}
{\rm Relative \; Difference}(n_{eff}) = \frac{\big[\frac{\sigma_{Y_{t-1}}^2}{(\overline{Y}_{N,t-1})^2} + \frac{\sigma_{Y_t}^2}{(\overline{Y}_{N,t})^2}\big]} {\big[\hat{\rho}_{Y_t, R_t} \times \sqrt{\frac{N_t - n_t}{n_t}} \times \frac{\sigma_{Y_t}}{\overline{Y}_{N,t}} - \hat{\rho}_{Y_{t-1}, R_{t-1}} \times \sqrt{\frac{N_{t-1} - n_{t-1}}{n_{t-1}}} \times \frac{\sigma_{Y_{t-1}}}{\overline{Y}_{N,{t-1}}} \big]^2} \\
\times \frac{1}{\big[1 - \hat{\rho}_{Y_{t-1}, R_{t-1}} \times \sqrt{\frac{N_{t-1} - n_{t-1}}{n_{t-1}}} \times \frac{\sigma_{Y_{t-1}}}{\overline{Y}_{N,{t-1}}} \big]^2}.    
\end{align*}

\subsection{Subgroup difference in means}
In this section, we present a method for evaluating the effective sample size for estimating mean differences between two subgroups following the decomposition approach in the study by \cite{Meng2018}. Researchers have noted the disparities in COVID-19 vaccine uptake between males and females using both the national level administrative data and cross-sectional surveys \citep{Hall2021-wo, Nassiri-Ansari2022-pp}. Let $\overline{Y}_{n_g}$ be the survey average of the variable $Y$ in the subgroup $g$, and let $\overline{Y}_{N_g}$ be the population average in the same subgroup. Let $n_g$ be the sample size of the survey conducted in the subgroup $g$, and let $N_g$ be the population size of the subgroup $g$. For simplicity, we assume the general population can be explicitly divided into two partitions: group I and group II, $\rm g = I, II$. To differentiate between these groups, we have created a binary indicator, represented by the variable $G$. If a response belongs to group II, the value of $G$ will be 1. If a response belongs to group I, the value of $G$ will be 0. Define a new variable $Y^{*}$, that is a combination of the variable $Y$ and the indicator $G$, $Y^{*} = Y \times G - Y \times (1-G)$. The revised formula for the decomposition of the estimation error can be written as follows: 

\begin{align}
    (\overline{Y}_{n_{\rm II}} - \overline{Y}_{n_{\rm I}}) - (\overline{Y}_{N_{\rm II}} - \overline{Y}_{N_{\rm I}}) = {\rho}_{Y^{*}, R^{*}} \times \sqrt{\frac{(N_{\rm I} + N_{\rm II}) - (n_{\rm I} + n_{\rm II})}{n_{\rm I} + n_{\rm II}}} \times \sigma_{Y^{*}},
\end{align}

\noindent where $R^{*}$ is a binary variable indicating if $Y^{*}$ is recorded. The focus of our study is to investigate the difference in vaccination rates between genders. In this context, we denote the estimated vaccination rate among females from the CTIS as $\overline{Y}_{n_{\rm II}}$, and the estimated vaccination rate among males from the CTIS as $\overline{Y}_{n_{\rm I}}$. The calculation of $\sigma_{Y^{*}}$ is presented in Appendix 1.5. We can then estimate ${\rho}_{Y^{*}, R^{*}}$ by substituting the known values from benchmark data into equation (5) and solving it. The interpretation of each factor is the same as that in the original formula. 
\\
\\
The revised formula for calculating the bias-adjusted effective sample size can be written as follows: 

\begin{align}
    n_{eff} = \frac{n_{\rm I} + n_{\rm II}}{(N_{\rm I} + N_{\rm II}) - (n_{\rm I} + n_{\rm II})} \times \frac{1}{(\hat{\rho}_{Y^{*}, R^{*}})^2}.
\end{align}

\subsection{Use as auxiliary information}

In this section, we use the decomposition framework for assessing the estimation accuracy of variable $Y$ with the assistance of auxiliary variable $X$. \cite{Deville1992} proposed the linear generalized regression estimator using a model between the variable of interest and the auxiliary variable. This method was subsequently extended to situations involving non-linear relationships between variables \citep{Firth1998}. Let $s$ be the notation of the survey sample among the finite population. Denote $N_t$ as the population size during time $t$. Assuming a constant relationship between the survey averages of $Y$ and $X$ throughout the study, we can fit a model between these two variables. By employing the auxiliary variable, we can generate estimates for the main variable, $\hat{y}_i$. Using these estimates, we can further provide a model-assisted survey estimate $\overline{Y}_{n^{*},t}$. The formula can be written as:

\begin{align}
    \overline{Y}_{n^{*},t} = \frac{\sum_{i \in s} y_i + \sum_{i \not\in s} \hat{y}_i}{N_t}.
\end{align}

\noindent For example, suppose we are interested in the model-assisted estimates of vaccine hesitancy among the US adult population ($Y$), taking into account the auxiliary information on vaccine uptake ($X$). While the CDC benchmark data serves as a reliable standard for vaccination uptake ($X$), there is no comparable benchmark available for assessing vaccine hesitancy ($Y$). Association between these two variables is widely reported in the literature, where vaccine hesitancy is a key hindrance in advancing optimal vaccine coverage among the population \citep{Chutiyami2022-se, Rane2022-ds}. Let the vaccine hesitancy $Y$ equal 1 if the respondent is unwilling to take the vaccine and equal 0 otherwise. Let the auxiliary variable $X$ be vaccine uptake. We assume that the entire population size $N_t$ stays constant throughout the study. Both the US CTIS and Axios-Ipsos survey inquired about vaccine uptake and vaccine hesitancy simultaneously during the study period. Given that our outcome, the sample average of vaccine hesitancy, is a proportion between 0 and 1, we use a beta regression model with the sample average of vaccine hesitancy and vaccine uptake from the Axios-Ipsos survey to estimate the relationship between these two variables \citep{ferrari2004beta}. Denote the sample average of vaccine hesitancy at time $t$ as $a_t$, and denote the sample average of vaccine uptake at time $t$ as $b_t$. The equation is as follows:

\begin{align}
    \log \big( \frac{\mu_t}{1 - \mu_t} \big) = \beta_0 + \beta_1 a_t, b_t \sim B(\mu_t, \phi).
\end{align}

\noindent Using the parameter estimates obtained from the model, we can calculate the model-assisted estimate $\overline{Y}_{n^{*},t}$ for both the Axios-Ipsos and CTIS. It is worth noting that we require the benchmark for vaccine uptake $X$ rather than vaccine hesitancy $Y$.
\\
\\
To apply the error decomposition framework in means, as proposed by \cite{Meng2018}, we make the assumption that the model-assisted estimate $\overline{Y}_{n^{*},t}$ obtained from the Axios-Ipsos survey serves as the benchmark for measuring vaccine hesitancy among US adults. We can then compare the original estimate $\overline{Y}_{n,t}$ with the model-assisted estimate $\overline{Y}_{n^{*},t}$ from the CTIS to decompose the components of the estimation error and the effective sample size. Similarly, we performed an adjustment for uncertainty by scaling the Axios-Ipsos survey's estimates by a factor of 0.9, 0.95, 1, 1.05, and 1.1.

\pagebreak

\section{Table and Figure}

\begin{table}[H]
\centering
\begin{tabular}{|c|cc|cc|c|}
\hline
\multirow{2}{*}{} &
  \multicolumn{2}{c|}{\textbf{CTIS}} &
  \multicolumn{2}{c|}{\textbf{CVoter survey}} &
  \multirow{2}{*}{\textbf{Census (2011)}} \\ \cline{2-5}
 &
  \multicolumn{1}{c|}{\textbf{Unweighted}} &
  \textbf{Weighted} &
  \multicolumn{1}{c|}{\textbf{Unweighted}} &
  \textbf{Weighted} &
   \\ \hline
\textbf{Gender}    & \multicolumn{1}{c|}{}     &      & \multicolumn{1}{c|}{}     &      &      \\ \hline
Female             & \multicolumn{1}{c|}{16\%} & 34\% & \multicolumn{1}{c|}{39\%} & 48\% & 48\% \\ \hline
Male               & \multicolumn{1}{c|}{84\%} & 66\% & \multicolumn{1}{c|}{61\%} & 52\% & 52\% \\ \hline
\textbf{Age}       & \multicolumn{1}{c|}{}     &      & \multicolumn{1}{c|}{}     &      &      \\ \hline
18-24 years        & \multicolumn{1}{c|}{23\%} & 21\% & \multicolumn{1}{c|}{19\%} & 22\% & 22\% \\ \hline
25-44 years        & \multicolumn{1}{c|}{63\%} & 53\% & \multicolumn{1}{c|}{54\%} & 47\% & 47\% \\ \hline
Above 45           & \multicolumn{1}{c|}{14\%} & 26\% & \multicolumn{1}{c|}{27\%} & 31\% & 31\% \\ \hline
\textbf{Education} & \multicolumn{1}{c|}{}     &      & \multicolumn{1}{c|}{}     &      &      \\ \hline
Less than college  & \multicolumn{1}{c|}{26\%} & 26\% & \multicolumn{1}{c|}{86\%} & 87\% & 87\% \\ \hline
College and above  & \multicolumn{1}{c|}{74\%} & 74\% & \multicolumn{1}{c|}{14\%} & 13\% & 13\% \\ \hline
\textbf{Location}  & \multicolumn{1}{c|}{}     &      & \multicolumn{1}{c|}{}     &      &      \\ \hline
Rural              & \multicolumn{1}{c|}{27\%} & 25\% & \multicolumn{1}{c|}{60\%} & 70\% & 69\% \\ \hline
Urban              & \multicolumn{1}{c|}{73\%} & 75\% & \multicolumn{1}{c|}{40\%} & 30\% & 31\% \\ \hline
\end{tabular} \\
\captionsetup{labelformat=empty, singlelinecheck = off}
\caption{\textbf{Table 1. Composition of the survey respondents by demographic variables in India.} The demographic characteristics of the respondents from the CTIS and the CVoter survey were compared to the latest Census (2011) in India. The CTIS included respondents in India from April 23, 2020, to July 10, 2021 (n = 2,243,662).}
\end{table}

\begin{table}[H]
\centering
\begin{tabular}{|c|ccc|ccc|}
\hline
\multirow{2}{*}{}  & \multicolumn{3}{c|}{\textbf{India}}                          & \multicolumn{3}{c|}{\textbf{US}}                             \\ \cline{2-7} 
 &
  \multicolumn{1}{c|}{\textbf{CTIS}} &
  \multicolumn{1}{c|}{\textbf{\begin{tabular}[c]{@{}c@{}}Facebook \\ active \\ users \\ (2021)\end{tabular}}} &
  \textbf{\begin{tabular}[c]{@{}c@{}}Census\\    (2011)\end{tabular}} &
  \multicolumn{1}{c|}{\textbf{CTIS}} &
  \multicolumn{1}{c|}{\textbf{\begin{tabular}[c]{@{}c@{}}Facebook \\ active \\ users \\ (2021)\end{tabular}}} &
  \textbf{\begin{tabular}[c]{@{}c@{}}Census\\    (2019)\end{tabular}} \\ \hline
\textbf{Gender}    & \multicolumn{1}{c|}{}     & \multicolumn{1}{c|}{}     &      & \multicolumn{1}{c|}{}     & \multicolumn{1}{c|}{}     &      \\ \hline
Female             & \multicolumn{1}{c|}{34\%} & \multicolumn{1}{c|}{24\%} & 48\% & \multicolumn{1}{c|}{52\%} & \multicolumn{1}{c|}{56\%} & 51\% \\ \hline
Male               & \multicolumn{1}{c|}{66\%} & \multicolumn{1}{c|}{76\%} & 52\% & \multicolumn{1}{c|}{46\%} & \multicolumn{1}{c|}{44\%} & 49\% \\ \hline
\textbf{Age}       & \multicolumn{1}{c|}{}     & \multicolumn{1}{c|}{}     &      & \multicolumn{1}{c|}{}     & \multicolumn{1}{c|}{}     &      \\ \hline
18-24 years old    & \multicolumn{1}{c|}{21\%} & \multicolumn{1}{c|}{36\%} & 22\% & \multicolumn{1}{c|}{11\%} & \multicolumn{1}{c|}{20\%} & 12\% \\ \hline
25-45 years old    & \multicolumn{1}{c|}{53\%} & \multicolumn{1}{c|}{54\%} & 47\% & \multicolumn{1}{c|}{33\%} & \multicolumn{1}{c|}{44\%} & 34\% \\ \hline
Above 45           & \multicolumn{1}{c|}{26\%} & \multicolumn{1}{c|}{11\%} & 31\% & \multicolumn{1}{c|}{56\%} & \multicolumn{1}{c|}{36\%} & 54\% \\ \hline
\textbf{Education} & \multicolumn{1}{c|}{}     & \multicolumn{1}{c|}{}     &      & \multicolumn{1}{c|}{}     & \multicolumn{1}{c|}{}     &      \\ \hline
Less than college  & \multicolumn{1}{c|}{26\%} & \multicolumn{1}{c|}{48\%} & 87\% & \multicolumn{1}{c|}{21\%} & \multicolumn{1}{c|}{44\%} & 39\% \\ \hline
College and above  & \multicolumn{1}{c|}{74\%} & \multicolumn{1}{c|}{52\%} & 13\% & \multicolumn{1}{c|}{79\%} & \multicolumn{1}{c|}{56\%} & 61\% \\ \hline
\textbf{\begin{tabular}[c]{@{}c@{}}Proportion in \\ total population\end{tabular}} &
  \multicolumn{1}{c|}{} &
  \multicolumn{1}{c|}{40\%} &
   &
  \multicolumn{1}{c|}{} &
  \multicolumn{1}{c|}{87\%} &
   \\ \hline
\end{tabular} \\
\captionsetup{labelformat=empty, singlelinecheck = off}
\caption{\textbf{Table 2. Composition of the sampling frames and weighted respondents from the CTIS in India and the US.} The CTIS included weighted respondents in India from April 23, 2020, to July 10, 2021 (n = 2,243,662), and weighted respondents in the US from April 6, 2020, to April 26, 2021 (n = 20,937,905). The latest Census (2011) and the March 2019 CPS
Supplements were used for benchmarks in India and the US, separately.}
\end{table}

\begin{table}[H]
\centering
\begin{tabular}{|c|cc|cc|}
\hline
\multirow{2}{*}{} &
  \multicolumn{2}{c|}{\textbf{\begin{tabular}[c]{@{}c@{}}India  \\ (May 16 to Sept. 18,   2021)\end{tabular}}} &
  \multicolumn{2}{c|}{\textbf{\begin{tabular}[c]{@{}c@{}}US   \\ (Feb. 7 to May 15,   2021)\end{tabular}}} \\ \cline{2-5} 
 &
  \multicolumn{1}{c|}{\textbf{CTIS}} &
  \textbf{\begin{tabular}[c]{@{}c@{}}CVoter\\ survey\end{tabular}} &
  \multicolumn{1}{c|}{\textbf{CTIS}} &
  \textbf{\begin{tabular}[c]{@{}c@{}}Axios-Ipsos\\ survey\end{tabular}} \\ \hline
\textbf{Benchmark (\%)} &
  \multicolumn{2}{c|}{\begin{tabular}[c]{@{}c@{}}33.24 \\ {[}15.08, 59.95{]}\end{tabular}} &
  \multicolumn{2}{c|}{\begin{tabular}[c]{@{}c@{}}40.07\\    {[}15.58, 61.91{]}\end{tabular}} \\ \hline
\textbf{Estimate (\%)} &
  \multicolumn{1}{c|}{\begin{tabular}[c]{@{}c@{}}75.40\\ {[}39.54, 89.82{]}\end{tabular}} &
  \begin{tabular}[c]{@{}c@{}}50.02\\ {[}21.30, 74.70{]}\end{tabular} &
  \multicolumn{1}{c|}{\begin{tabular}[c]{@{}c@{}}52.92\\ {[}20.72, 76.91{]}\end{tabular}} &
  \begin{tabular}[c]{@{}c@{}}30.50\\ {[}15.00, 64.00{]}\end{tabular} \\ \hline
\textbf{Estimation error} &
  \multicolumn{1}{c|}{\begin{tabular}[c]{@{}c@{}}0.39\\ {[}0.24, 0.43{]}\end{tabular}} &
  \begin{tabular}[c]{@{}c@{}}0.16\\ {[}0.06, 0.20{]}\end{tabular} &
  \multicolumn{1}{c|}{\begin{tabular}[c]{@{}c@{}}0.13\\ {[}0.05, 0.16{]}\end{tabular}} &
  \begin{tabular}[c]{@{}c@{}}0.01\\ {[}-0.01, 0.04{]}\end{tabular} \\ \hline
\textbf{Data deficiency} &
  \multicolumn{1}{c|}{\begin{tabular}[c]{@{}c@{}}224\\    {[}150, 242{]}\end{tabular}} &
  \begin{tabular}[c]{@{}c@{}}588\\    {[}481, 943{]}\end{tabular} &
  \multicolumn{1}{c|}{\begin{tabular}[c]{@{}c@{}}33\\    {[}31, 39{]}\end{tabular}} &
  \begin{tabular}[c]{@{}c@{}}535\\    {[}510, 554{]}\end{tabular} \\ \hline
\textbf{Problem difficulty} &
  \multicolumn{2}{c|}{\begin{tabular}[c]{@{}c@{}}0.47\\ {[}0.36, 0.50{]}\end{tabular}} &
  \multicolumn{2}{c|}{\begin{tabular}[c]{@{}c@{}}0.49\\    {[}0.37, 0.50{]}\end{tabular}} \\ \hline
\textbf{\begin{tabular}[c]{@{}c@{}}Data defect \\ correlation\end{tabular}} &
  \multicolumn{1}{c|}{\begin{tabular}[c]{@{}c@{}}0.0040\\    {[}0.0027, 0.0046{]}\end{tabular}} &
  \begin{tabular}[c]{@{}c@{}}0.0005\\    {[}0.0003, 0.0007{]}\end{tabular} &
  \multicolumn{1}{c|}{\begin{tabular}[c]{@{}c@{}}0.0079\\    {[}0.0045, 0.0091{]}\end{tabular}} &
  \begin{tabular}[c]{@{}c@{}}0.0001\\    {[}-0.0001, 0.0002{]}\end{tabular} \\ \hline
\textbf{\begin{tabular}[c]{@{}c@{}}Effective \\ sample size \end{tabular}} &
  \multicolumn{1}{c|}{\begin{tabular}[c]{@{}c@{}}2\\    {[}1, 3{]}\end{tabular}} &
  \begin{tabular}[c]{@{}c@{}}9\\    {[}6, 33{]}\end{tabular} &
  \multicolumn{1}{c|}{\begin{tabular}[c]{@{}c@{}}15\\    {[}10, 50{]}\end{tabular}} &
  \begin{tabular}[c]{@{}c@{}}859\\    {[}138, 980{]}\end{tabular} \\ \hline
\end{tabular} \\
\captionsetup{labelformat=empty, singlelinecheck = off}
\caption{\textbf{Table 3. Estimation errors and their components of vaccine uptake from surveys in India and the US.} The benchmark data for India and the US were from the CoWIN and the CDC, separately. Each continuous variable reported its median, minimum, and maximum in the form of the median [minimum, maximum].}
\end{table}

\begin{table}[H]
\centering
\begin{tabular}{|c|cc|}
\hline
\multirow{2}{*}{\textbf{}} &
  \multicolumn{2}{c|}{\textbf{India}} \\ \cline{2-3} 
 &
  \multicolumn{1}{c|}{\textbf{\href{https://covidmap.umd.edu/}{CTIS}}} &
  \textbf{\href{https://cvoterindia.com/trackers/}{CVoter survey}} \\ \hline
\textbf{Survey type} &
  \multicolumn{1}{c|}{Non-probability survey} &
  Probability survey \\ \hline
\textbf{Target population} &
  \multicolumn{1}{c|}{Indian adults} &
  Indian adults \\ \hline
\textbf{Sampling frame} &
  \multicolumn{1}{c|}{\begin{tabular}[c]{@{}c@{}}Adult Facebook \\ Active User Base\end{tabular}} &
  \begin{tabular}[c]{@{}c@{}}Adult telephone \\ subscribers\end{tabular} \\ \hline
\textbf{Recruitment mode} &
  \multicolumn{1}{c|}{Facebook newsfeed} &
  \begin{tabular}[c]{@{}c@{}}Computer-assisted \\ telephone interview\end{tabular} \\ \hline
\textbf{Interview mode} &
  \multicolumn{1}{c|}{Online} &
  Telephone \\ \hline
\textbf{\begin{tabular}[c]{@{}c@{}}Average sample size \\ per week\end{tabular}} &
  \multicolumn{1}{c|}{25,000} &
  2,700 \\ \hline
\textbf{Response rate} &
  \multicolumn{1}{c|}{1\%} &
  55\% \\ \hline
\textbf{\begin{tabular}[c]{@{}c@{}}Vaccine uptake \\ question\end{tabular}} &
  \multicolumn{1}{c|}{\begin{tabular}[c]{@{}c@{}}“Have you had a \\ COVID-19 vaccination?”\end{tabular}} &
  \begin{tabular}[c]{@{}c@{}}“Have you got your \\ COVID-19 vaccine shots?”\end{tabular} \\ \hline
\textbf{\begin{tabular}[c]{@{}c@{}}Vaccine uptake \\ responses\end{tabular}} &
  \multicolumn{1}{c|}{“Yes”} &
  \begin{tabular}[c]{@{}c@{}}“Yes, have got one shot/\\ both shots of the vaccine”\end{tabular} \\ \hline
\textbf{\begin{tabular}[c]{@{}c@{}}Vaccine hesitancy \\ question\end{tabular}} &
  \multicolumn{1}{c|}{\begin{tabular}[c]{@{}c@{}}“If a vaccine to prevent \\ COVID-19 were offered to \\ you today, would you choose \\ to get vaccinated?”\end{tabular}} &
  \begin{tabular}[c]{@{}c@{}}“When a new Coronavirus \\ (COVID-19) vaccine becomes \\ publicly available you \\ would take it?”\end{tabular} \\ \hline
\textbf{\begin{tabular}[c]{@{}c@{}}Vaccine hesitancy \\ responses\end{tabular}} &
  \multicolumn{1}{c|}{“No, probably/definitely not”} &
  “Disagree/strongly disagree” \\ \hline
\textbf{\begin{tabular}[c]{@{}c@{}}Weighting \\ variables\end{tabular}} &
  \multicolumn{1}{c|}{Age, gender, region} &
  \begin{tabular}[c]{@{}c@{}}Age, gender, education, \\ income, social group, \\ rurality, and region\end{tabular} \\ \hline
\textbf{\begin{tabular}[c]{@{}c@{}}Sources for \\ demographic \\ benchmarks\end{tabular}} &
  \multicolumn{1}{c|}{\begin{tabular}[c]{@{}c@{}}UN Population Division 2019 \\ World Population Projection\end{tabular}} &
  \begin{tabular}[c]{@{}c@{}}The latest census (2011) and \\ national sample surveys \\ (NSS) estimates\end{tabular} \\ \hline
\end{tabular} \\
\captionsetup{labelformat=empty, singlelinecheck = off}
\caption{\textbf{Table 4. Comparisons of the survey designs in India.} The benchmark data for vaccine uptake were sourced from the \href{https://dashboard.cowin.gov.in/}{CoWIN}.}
\end{table}

\begin{figure}[H]
    \centering
    \includegraphics[width = 1\linewidth]{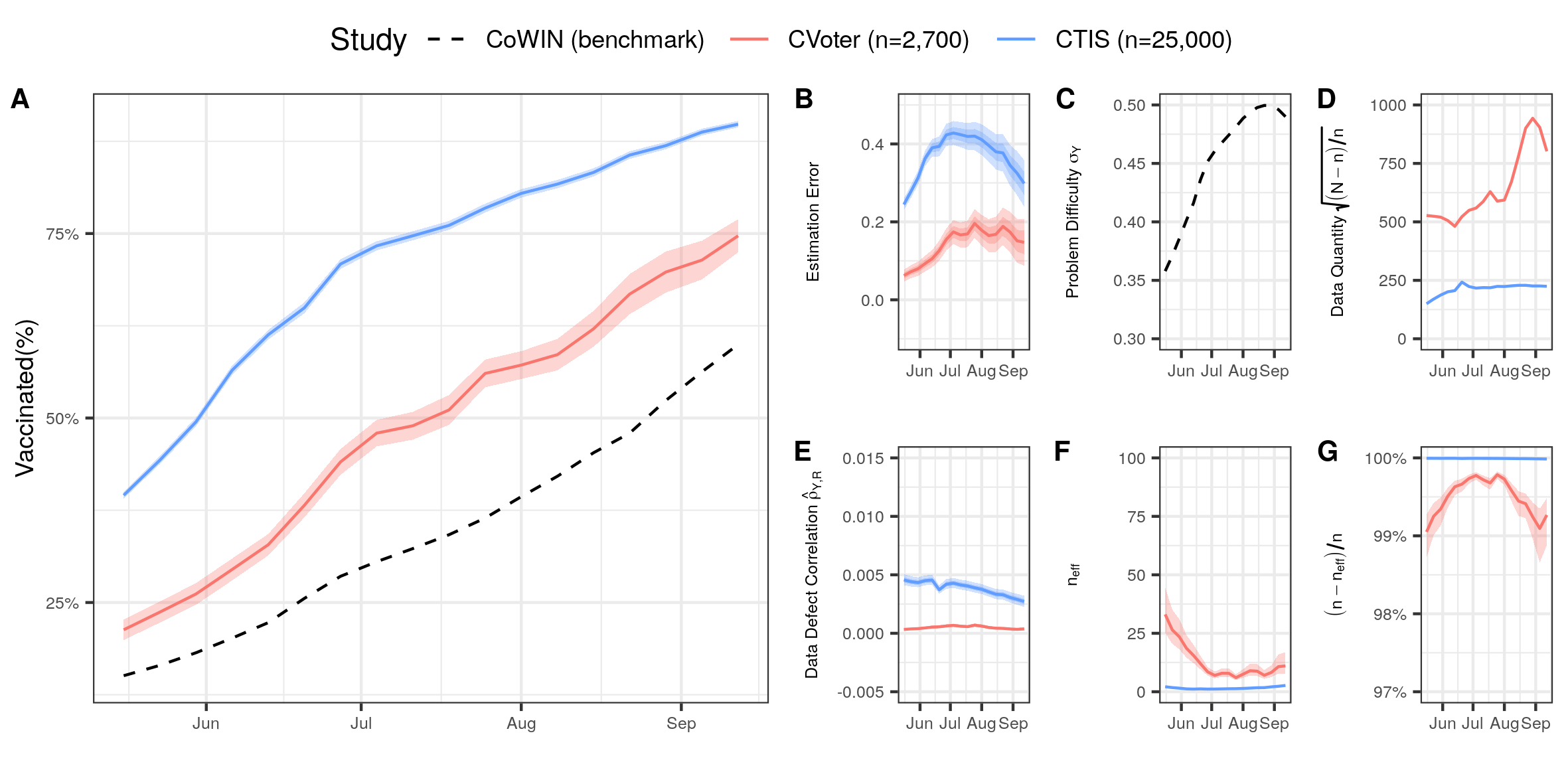}
    \captionsetup{labelformat=empty, singlelinecheck = off}
    \caption{\textbf{Figure 1. Decomposition of the estimation error and the effective sample size of vaccine uptake in India.} (A) Estimate, (B) estimation error, (c) inherent problem difficulty, (D) data deficiency, (E) data defect correlation, (F) effective sample size, and (G) reduction in the sample size of vaccine uptake among Indian adults from the CTIS (blue) and the CVoter survey (red) compared to the CoWIN benchmark (black) between May 16 and September 18, 2021. For plot A, shaded bands show the classic 95\% confidence intervals of the estimates. For plots (B) and (E), shaded bands show the $\pm$ 5\% and $\pm$ 10\% benchmark imprecision adjustments. For plots (F) and (G), shaded bands show the $\pm$ 5\% benchmark imprecision adjustments.}
\end{figure}

\begin{figure}[H]
    \centering
    \includegraphics[width = 1\linewidth]{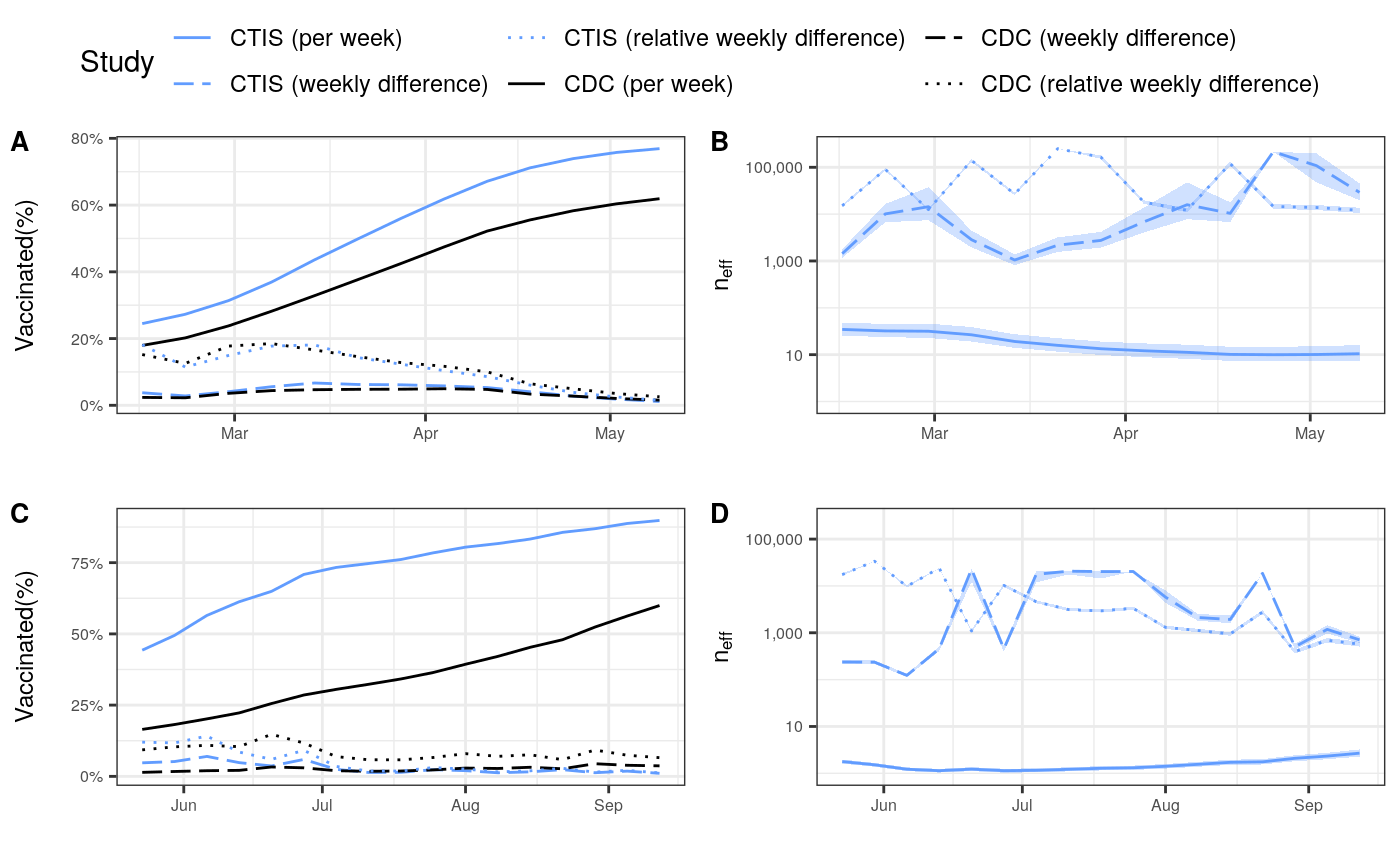}
    \captionsetup{labelformat=empty, singlelinecheck = off}
    \caption{\textbf{Figure 2. The successive difference and relative successive difference of vaccine uptake in the US ((A) and (B)) and India ((C) and (D)).} (A) The estimate and (B) effective sample size of the overall rate (blue, solid), the successive difference (blue, dashed), and the relative successive difference (blue, dotted) of vaccine uptake among US adults from the CTIS compared to the CDC benchmark (black, solid; black, dashed; and black, dotted) between February 7 and May 15, 2021. (C) The estimate and (D) effective sample size of the overall rate (blue, solid), the successive difference (blue, dashed), and the relative successive difference (blue, dotted) of vaccine uptake among Indian adults from the CTIS compared to the CoWIN benchmark (black, solid; black, dashed; and black, dotted) between May 16 and September 18, 2021. For plots (B) and (D), shaded bands show the $\pm$ 5\% benchmark imprecision adjustments.}
\end{figure}

\begin{figure}[H]
    \centering
    \includegraphics[width = 1\linewidth]{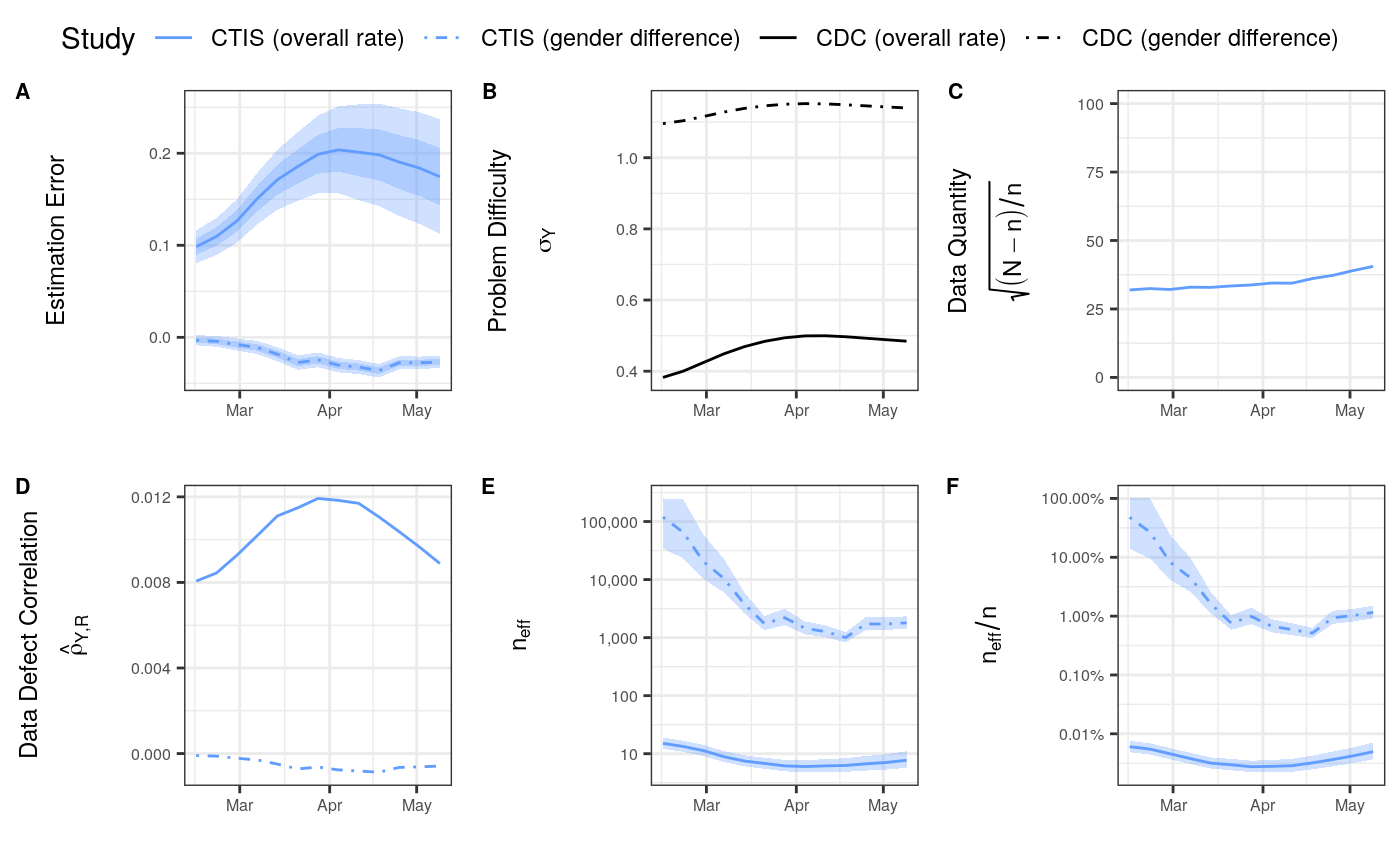}
    \captionsetup{labelformat=empty, singlelinecheck = off}
    \caption{\textbf{Figure 3. Decomposition of the estimation error and effective sample size of gender difference in vaccine uptake among US adults.} (A) Estimation error, (B) inherent problem difficulty, (C) data deficiency, (D) data defect correlation, (E) effective sample size, and (F) its proportion in the sample size of the overall rate (solid) and the gender difference (dot-dash) of vaccine uptake among US adults from the CTIS (blue) compared to the CDC benchmark (black) between February 7 and May 16, 2021. For plots (A) and (D), shaded bands show the $\pm$ 5\% and $\pm$ 10\% benchmark imprecision adjustments. For plots (E) and (F), shaded bands show the $\pm$ 5\% benchmark imprecision adjustments.}
\end{figure}

\begin{figure}[H]
    \centering
    \includegraphics[width = 1\linewidth]{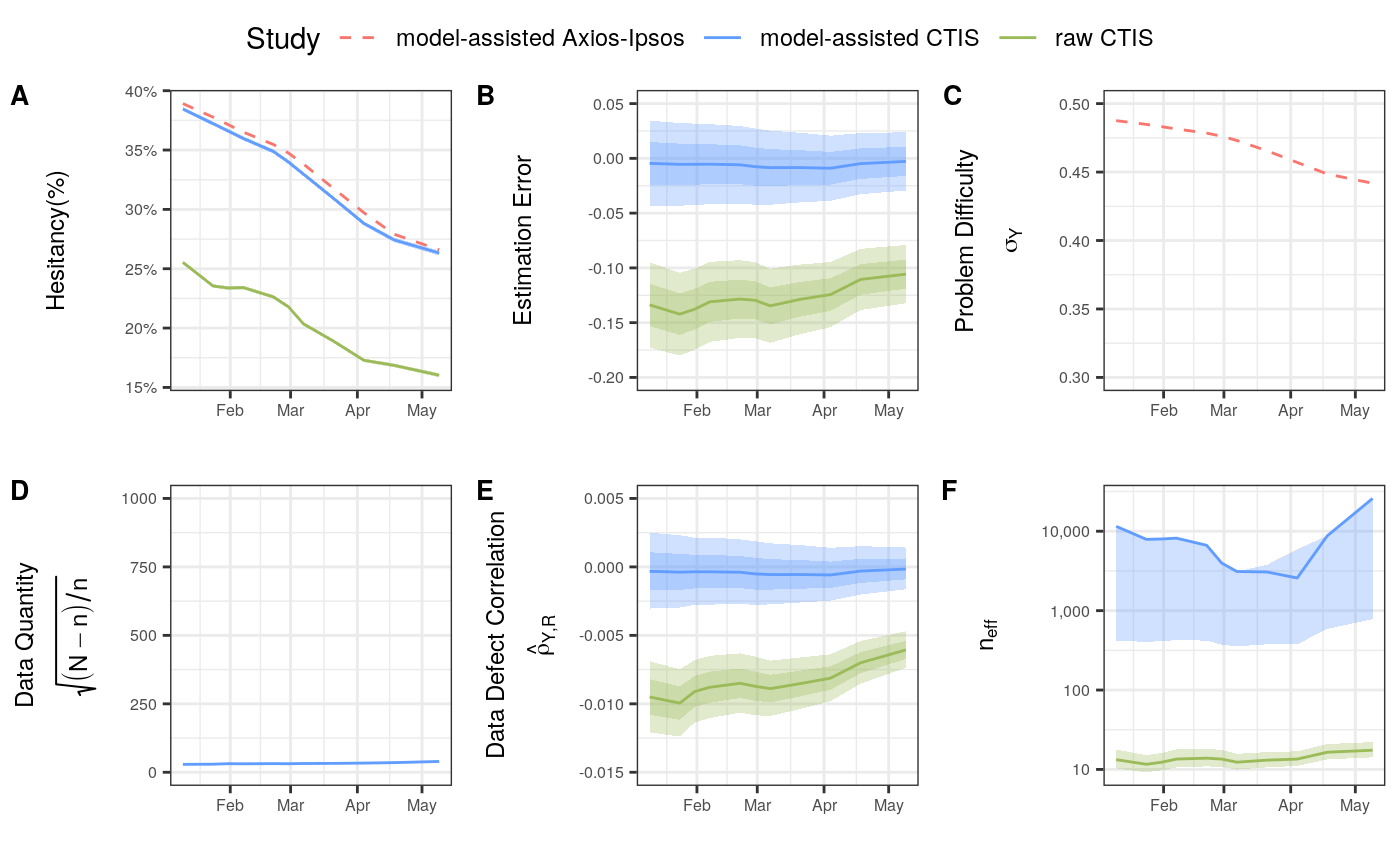}
    \captionsetup{labelformat=empty, singlelinecheck = off}
    \caption{\textbf{Figure 4. Decomposition of the estimation error and effective sample size of vaccine hesitancy in the US.} (A) The estimate, (B) estimation error, (C) inherent problem difficulty, (D) data deficiency, (E) data defect correlation, and (F) effective sample size of the vaccine hesitancy among US adults from the model-assisted CTIS (blue) and original CTIS (green), compared to the model-assisted Axios-Ipsos benchmark (red) between January 10 and May 15, 2021. For plots (B) and (E), shaded bands show the $\pm$ 5\% and $\pm$ 10\% benchmark imprecision adjustments. For plot (F), shaded bands show the $\pm$ 5\% benchmark imprecision adjustments.}
\end{figure}

\pagebreak

\section{Acknowledgments}

We would like to express our sincere gratitude to the Social Data Science Center at the University of Maryland and the Delphi Group at Carnegie Mellon University for their invaluable contribution to conducting the CTIS and providing us access to individual-level data. We would also like to extend our thanks to Professor F. Kreuter for their valuable insights and to Professor X.-L. Meng for their inspired discussions on this topic. The authors would like to thank Lauren Zimmermann for her help in data processing.

\subsection{Funding}

The study was supported by NSF DMS 1712933, a pilot award from the Michigan Institute of Data Science and the University of Michigan Rogel Cancer Center.

\subsection{Author contributions}

The authors confirm their contribution to the paper as follows: Conceptualization: B.M. and W.D. Methodology: Y.Y., W.D., P.H., and B.M. Software: Y.Y. Formal analysis: Y.Y., W.D., P.H., and B.M. Resources: Y.D.  and B.M. Writing (original draft preparation): Y.Y., W.D., P.H., and B.M. Writing (review and editing): Y.Y., W.D., P.H., Y.D., S.R., B.T., and B.M. Visualization: Y.Y. and B.M. Supervision: B.M. Funding acquisition: B.M. All authors reviewed the results and approved the final version of the manuscript. 

\subsection{Competing interests}

The authors declare that they have no competing interests.

\subsection{Data availability}

All data needed to evaluate the conclusions in the paper are present in the paper and/or the Supplementary Materials. The code for reproducing the findings can be found in the repository at https://github.com/youqiy/bigdataparadox.

\pagebreak

\section{Supplementary Texts}

\subsection{Effective sample size of the successive difference in means}

This section illustrates how to calculate the effective sample size of the successive difference in means, which is complementary to Section 4.4. We assume that the two simple random samplings are conducted at consecutive times with the same sample size and that the size of the finite population being studied $N$ remains constant over time. Furthermore, we assume that the observations from the target surveys conducted at consecutive times are independent of each other. The variance of two SRS surveys can be computed by the following equation:

\begin{align*}
    Var(\overline{Y}_{n_{eff}, t} - \overline{Y}_{n_{eff}, {t-1}}) = \frac{\sigma^2_{Y_{t-1}} + \sigma^2_{Y_{t}}}{N -1} \frac{N - n_{eff}}{n_{eff}}.
\end{align*}

\noindent By equating the squared term of the estimation error $[(\bar{Y}_{n,t} - \bar{Y}_{n,t-1}) - (\bar{Y}_{N,t} - \bar{Y}_{N,t-1})]$ to the above variance, it yields the following equation:

\begin{align*}
    n_{eff} = \frac{(\sigma^2_{Y_{t-1}} + \sigma^2_{Y_{t}}) \frac{N}{{N-1}}}{(\sigma^2_{Y_{t-1}} + \sigma^2_{Y_{t}}) \frac{1}{{N-1}} + [(\overline{Y}_{n,t} - \overline{Y}_{n,t-1})-(\overline{Y}_{N,t} - \overline{Y}_{N,t-1})]^2} \approx \frac{\sigma_{Y_{t-1}}^2 + \sigma_{Y_t}^2}{[(\overline{Y}_{n,t} - \overline{Y}_{n,t-1})-(\overline{Y}_{N,t} - \overline{Y}_{N,t-1})]^2}.
\end{align*}

\subsection{Effective sample size of the relative successive difference in means}

This section illustrates how to calculate the effective sample size of the relative successive difference in means, which is complementary to Section 4.4. Similarly, we assume equal sample sizes in two SRS at consecutive times, unchanged population size, and independent observations. Using the Taylor polynomial of a function of two variables, the variance of two SRS surveys can be calculated by: 

\begin{align*}
    Var\Big(\frac{\overline{Y}_{n_{eff}, t} - \overline{Y}_{n_{eff}, {t-1}}}{\overline{Y}_{n_{eff}, {t-1}}}\Big) = Var\Big(\frac{\overline{Y}_{n_{eff}, t}}{\overline{Y}_{n_{eff}, {t-1}}}\Big) \approx \frac{1}{n_{eff}} {\Big(\frac{\overline{Y}_{N,t}}{\overline{Y}_{N,t-1}}\Big)^2\big[\frac{\sigma_{Y_{t-1}}^2}{(\overline{Y}_{N,t-1})^2} + \frac{\sigma_{Y_t}^2}{(\overline{Y}_{N,t})^2}\big]}
\end{align*}

\noindent By equating the squared term of the estimation error $[(\overline{Y}_{n,t} - \overline{Y}_{n,t-1}) \slash \overline{Y}_{n,t-1} - (\overline{Y}_{N,t} - \overline{Y}_{N,t-1}) \slash \overline{Y}_{N,t-1}]$ to the above variance, it yields the following formula:

\begin{align*}
     n_{eff} = \frac{\Big(\frac{\overline{Y}_{N,t}}{\overline{Y}_{N,t-1}}\Big)^2\big[\frac{\sigma_{Y_{t-1}}^2}{(\overline{Y}_{N,t-1})^2} + \frac{\sigma_{Y_t}^2}{(\overline{Y}_{N,t})^2}\big]} {\big[\frac{\overline{Y}_{n,t} - \overline{Y}_{n,t-1}}{\overline{Y}_{n,t-1}} - \frac{\overline{Y}_{N,t} - \overline{Y}_{N,t-1}}{\overline{Y}_{N,t-1}}\big]^2}.
\end{align*}

\subsection{Reformulating the effective sample size of the successive difference in means}

This section illustrates how to represent the effective sample size of the successive difference in means in terms of the three components of estimation error, which is complementary to Section 4.4. In equation (3), the denominator represents the squared term of the estimation error associated with the successive difference. By utilizing the decomposition of error across different dates, we can express the estimation error of the successive difference in the following manner. 

\begin{align*}
    (\overline{Y}_{n,t} - \overline{Y}_{n,t-1})-(\overline{Y}_{N,t} - \overline{Y}_{N,t-1}) = (\overline{Y}_{n,t} - \overline{Y}_{N,t})-(\overline{Y}_{n,t-1} - \overline{Y}_{N,t-1}) \\ = \hat{\rho}_{Y_t, R_t} \times \sqrt{\frac{N_t - n_t}{n_t}} \times \sigma_{Y_t} -\hat{\rho}_{Y_{t-1}, R_{t-1}} \times \sqrt{\frac{N_{t-1} - n_{t-1}}{n_{t-1}}} \times \sigma_{Y_{t-1}}.
\end{align*}

\subsection{Reformulating the effective sample size of the relative successive difference in means}

This section illustrates how to represent the effective sample size of the relative successive difference in means in terms of the three components of estimation error, which is complementary to Section 4.4. Using the second-degree Taylor polynomial of a function of two variables, we can express the estimation error of the relative successive difference in the following manner. 

\begin{align*}
    \frac{\overline{Y}_{n,t} - \overline{Y}_{n,t-1}}{\overline{Y}_{n,t-1}} - \frac{\overline{Y}_{N,t} - \overline{Y}_{N,t-1}}{\overline{Y}_{N,t-1}} = \frac{\overline{Y}_{N,t}}{\overline{Y}_{N,t-1}} \\
    \times \left(\hat{\rho}_{Y_t, R_t} \times \sqrt{\frac{N_t - n_t}{n_t}} \times \frac{\sigma_{Y_t}}{\overline{Y}_{N,t}} - \hat{\rho}_{Y_{t-1}, R_{t-1}} \times \sqrt{\frac{N_{t-1} - n_{t-1}}{n_{t-1}}} \times \frac{\sigma_{Y_{t-1}}}{\overline{Y}_{N,{t-1}}} \right) \\
\times \left(1 - \hat{\rho}_{Y_{t-1}, R_{t-1}} \times \sqrt{\frac{N_{t-1} - n_{t-1}}{n_{t-1}}} \times \frac{\sigma_{Y_{t-1}}}{\overline{Y}_{N,{t-1}}} \right).
\end{align*}

\subsection{Variance for subgroup difference}

This section presents the method for computing the value of $\sigma^2_{Y^*}$ for the error decomposition of the subgroup difference, which is complementary to Section 4.5. In our study of gender differences in vaccine uptake, both $Y$ and $G$ follow a Bernoulli distribution. By using the property of the variance under the Bernoulli distribution, we can derive,

\begin{align*}
    Var(\sigma^2_{Y^*}) = \overline{Y}_N (1 - \overline{Y}_N) + 4 \overline{G}_N (1 - \overline{G}_N) + 4[E(Y = 1, G = 1) - \overline{Y}_N \times \overline{G}_N].
\end{align*}

\noindent Here, $\overline{Y}_N$ is the population mean of the vaccination rate, $\overline{G}_N$ is the proportion of females among US adults, and E(Y = 1, G = 1) is the proportion of vaccinated females in the general adult population.

\pagebreak

\section{Supplementary Tables and Figures}

\begin{table}[H]
\centering
\begin{tabular}{|c|cc|}
\hline
\multirow{2}{*}{\textbf{}} &
  \multicolumn{2}{c|}{\textbf{US}} \\ \cline{2-3} 
 &
  \multicolumn{1}{c|}{\textbf{\href{https://cmu-delphi.github.io/delphi-epidata/symptom-survey/}{CTIS}}} &
  \textbf{\href{https://www.ipsos.com/en-us/news-polls/axios-ipsos-coronavirus-index}{Axios-Ipsos survey}} \\ \hline
\textbf{Survey type} &
  \multicolumn{1}{c|}{Non-probability survey} &
  Probability survey \\ \hline
\textbf{Target population} &
  \multicolumn{1}{c|}{US adults} &
  US adults \\ \hline
\textbf{Sampling frame} &
  \multicolumn{1}{c|}{\begin{tabular}[c]{@{}c@{}}Adult Facebook \\ Active User Base\end{tabular}} &
  Ipsos KnowledgePanel \\ \hline
\textbf{Recruitment mode} &
  \multicolumn{1}{c|}{Facebook newsfeed} &
  \begin{tabular}[c]{@{}c@{}}Addressed-based \\ mail sample\end{tabular} \\ \hline
\textbf{Interview mode} &
  \multicolumn{1}{c|}{Online} &
  Online \\ \hline
\textbf{\begin{tabular}[c]{@{}c@{}}Average sample size \\ per week\end{tabular}} &
  \multicolumn{1}{c|}{250,000} &
  1,000 \\ \hline
\textbf{Response rate} &
  \multicolumn{1}{c|}{1\%} &
  50\% \\ \hline
\textbf{\begin{tabular}[c]{@{}c@{}}Vaccine uptake \\ question\end{tabular}} &
  \multicolumn{1}{c|}{\begin{tabular}[c]{@{}c@{}}“Have you had a \\ COVID-19 vaccination?”\end{tabular}} &
  \begin{tabular}[c]{@{}c@{}}“Do you personally know anyone \\ who has already received \\ the COVID-19 vaccine?”\end{tabular} \\ \hline
\textbf{\begin{tabular}[c]{@{}c@{}}Vaccine uptake \\ responses\end{tabular}} &
  \multicolumn{1}{c|}{“Yes”} &
  \begin{tabular}[c]{@{}c@{}}“Yes, I have \\ received the vaccine”\end{tabular} \\ \hline
\textbf{\begin{tabular}[c]{@{}c@{}}Vaccine hesitancy \\ question\end{tabular}} &
  \multicolumn{1}{c|}{\begin{tabular}[c]{@{}c@{}}“If a vaccine to prevent \\ COVID-19 were offered to \\ you today, would you choose \\ to get vaccinated?”\end{tabular}} &
  \begin{tabular}[c]{@{}c@{}}“How likely, if at all, \\ are you to get the first \\ generation COVID-19 \\ vaccine, as soon as \\ it’s available”\end{tabular} \\ \hline
\textbf{\begin{tabular}[c]{@{}c@{}}Vaccine hesitancy \\ responses\end{tabular}} &
  \multicolumn{1}{c|}{“No, probably/definitely not”} &
  “Not very/at all likely” \\ \hline
\textbf{\begin{tabular}[c]{@{}c@{}}Weighting \\ variables\end{tabular}} &
  \multicolumn{1}{c|}{Age, gender, region} &
  \begin{tabular}[c]{@{}c@{}}Age, gender, education, \\ Census region, income, \\ social group, urbanity, \\ partisanship, and region\end{tabular} \\ \hline
\textbf{\begin{tabular}[c]{@{}c@{}}Sources for \\ demographic \\ benchmarks\end{tabular}} &
  \multicolumn{1}{c|}{\begin{tabular}[c]{@{}c@{}}March 2017 CPS \\ Supplement\end{tabular}} &
  \begin{tabular}[c]{@{}c@{}}March 2019 CPS\\ Supplement\end{tabular} \\ \hline
\end{tabular} \\
\captionsetup{labelformat=empty, singlelinecheck = off}
\caption{\textbf{Table S1. Comparisons of the survey designs in the US.} The benchmark data for vaccine uptake were sourced from the \href{https://www.cdc.gov/coronavirus/2019-ncov/vaccines/reporting-vaccinations.html}{CDC} reports.}
\end{table}

\begin{figure}[H]
    \centering
    \includegraphics[width = 1\linewidth]{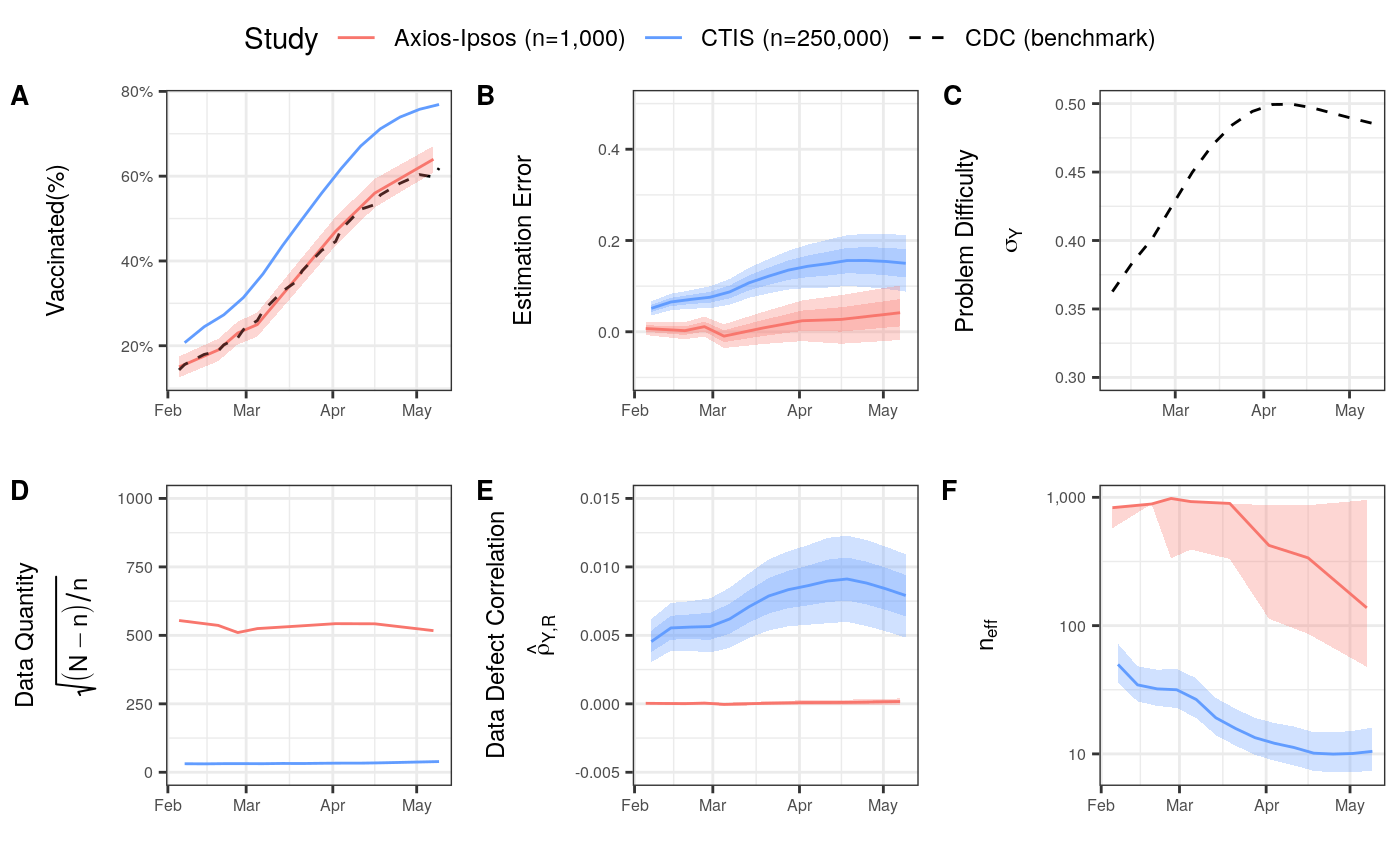}
    \captionsetup{labelformat=empty, singlelinecheck = off}
    \caption{\textbf{Figure S1. Decomposition of the estimation error of vaccine uptake in the US.} (A) The estimate, (B) estimation error, (C) inherent problem difficulty, (D) data deficiency, (E) data defect correlation, and (F) effective sample size of vaccine uptake among US adults from the CTIS (blue) and the Axios-Ipsos survey (red), compared to the CDC benchmark (black) between February 7 and May 15, 2021. For plot (A), shaded bands show the classic 95\% confidence interval. For plots (B) and (E), shaded bands show the $\pm$ 5\% and $\pm$ 10\% benchmark imprecision adjustments. For plot (F), shaded bands show the $\pm$ 5\% benchmark imprecision adjustments.}
\end{figure}

\begin{figure}[H]
    \centering
    \includegraphics[width = 0.8\linewidth]{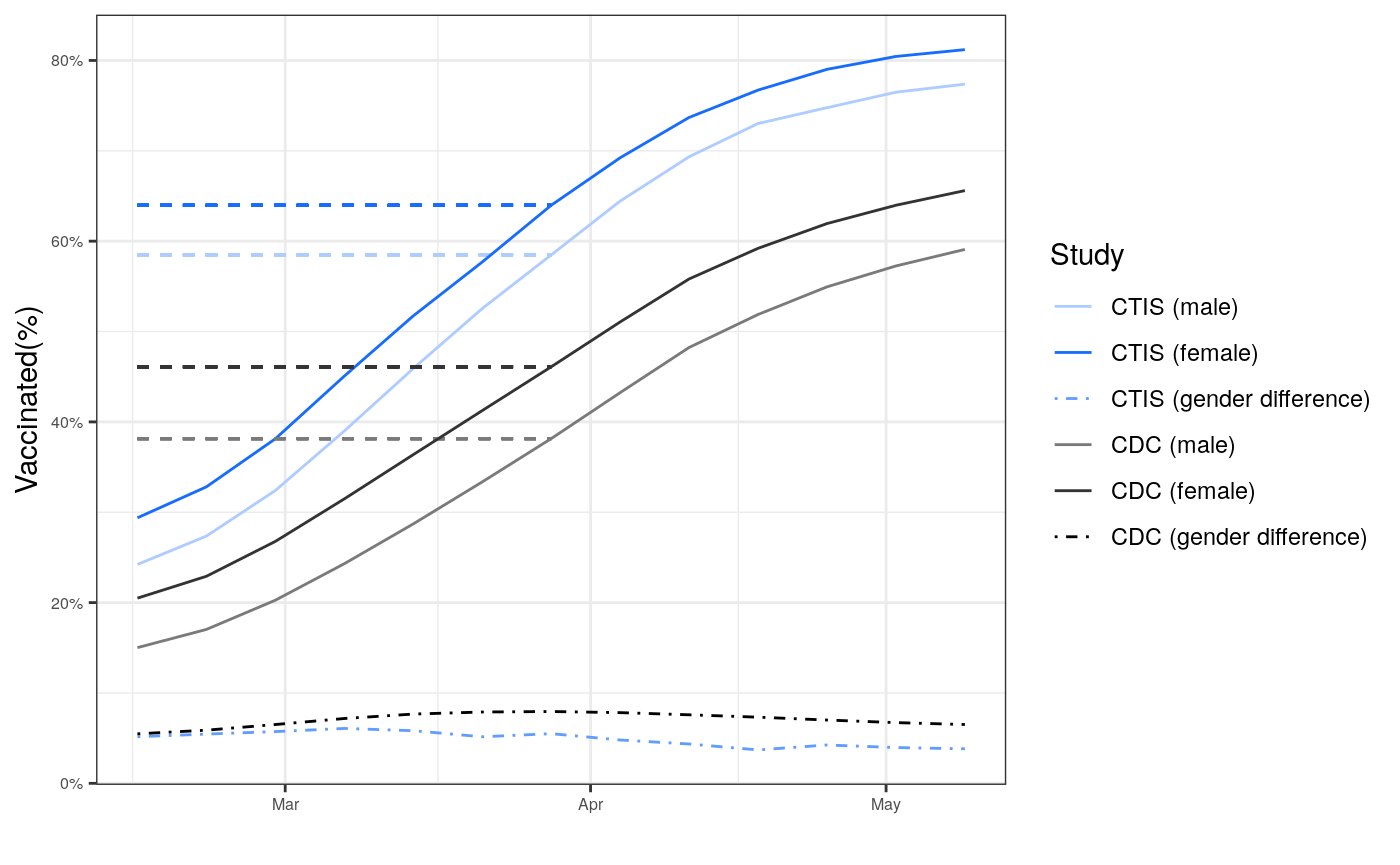}
    \captionsetup{labelformat=empty, singlelinecheck = off}
    \caption{\textbf{Figure S2. The difference in vaccine uptake between genders in the US.}  Estimates from the CTIS on vaccine uptake among US adults in males (light blue, solid), females (dark blue, solid), and the gender difference (blue, dot-dash) compared to the CDC benchmark (light gray, solid; dark gray, solid; black, dot-dash) between February 7 and May 15, 2021. The horizontal dashed lines indicate the median vaccination rate from each data regarding each gender.}
\end{figure}

\begin{figure}[H]
    \centering
    \includegraphics[width = 0.7\linewidth]{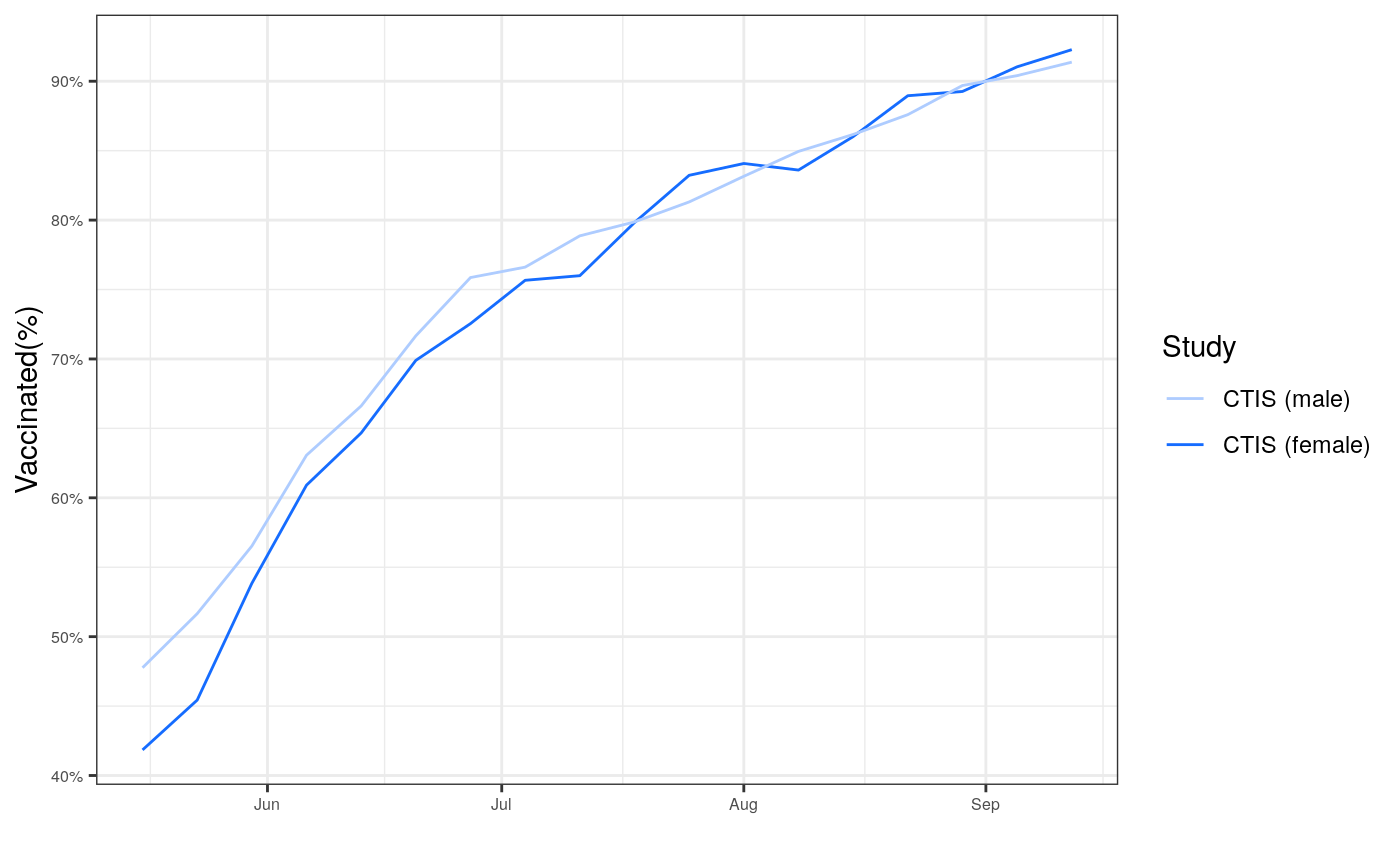}
    \captionsetup{labelformat=empty, singlelinecheck = off}
    \caption{\textbf{Figure S3. The difference in vaccine uptake between genders in India.} Estimates from the CTIS on vaccine uptake among Indian adults in males (light blue, solid) and females (dark blue, solid) between May 16 and September 18, 2021.}
\end{figure}

\begin{figure}[H]
    \centering
    \includegraphics[width = 0.8\linewidth]{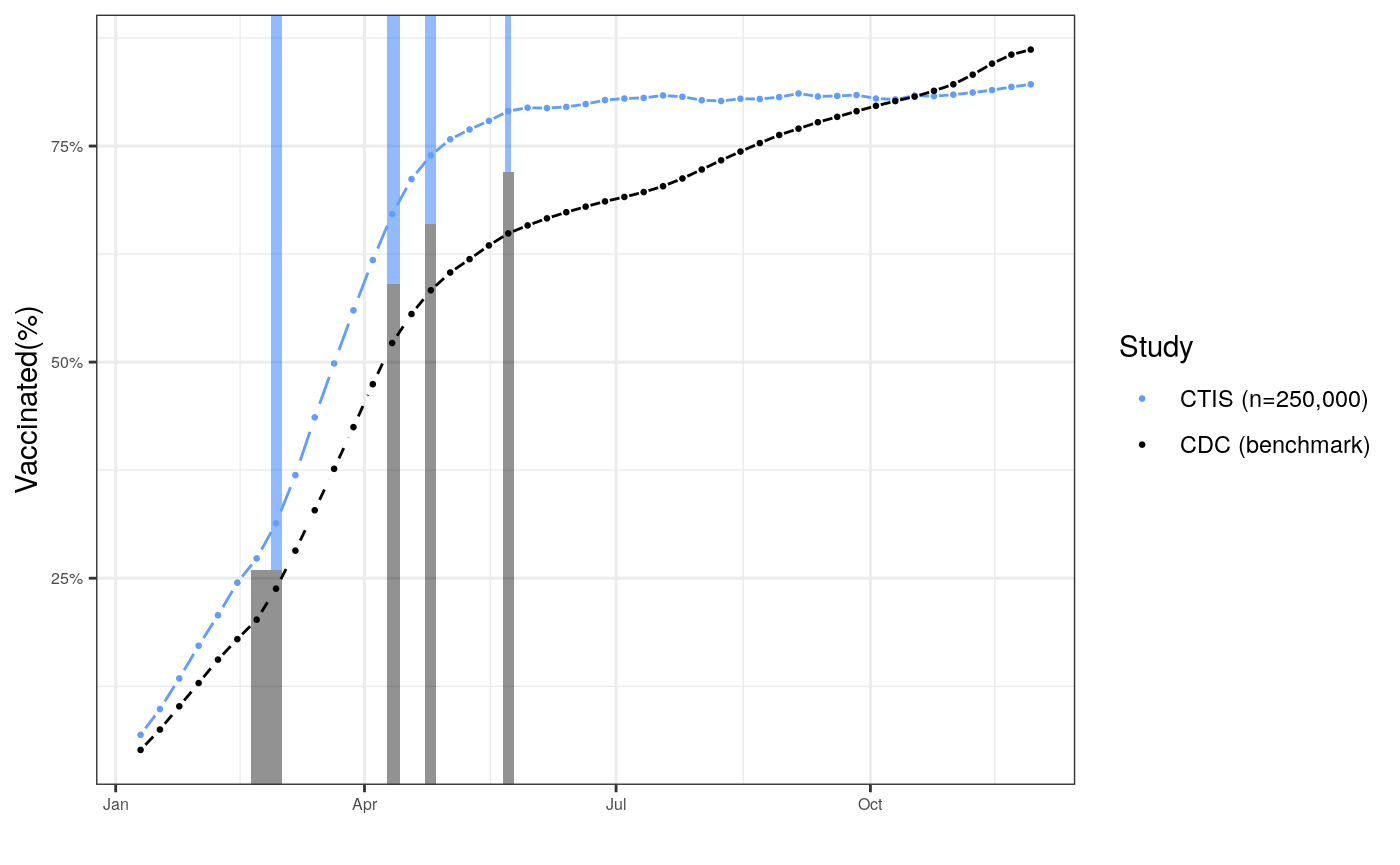}
    \captionsetup{labelformat=empty, singlelinecheck = off}
    \caption{\textbf{Figure S4. Change point detection in vaccine uptake among US adults.} Change points in the time series of the vaccination rates from the CTIS (blue) and the CDC benchmark (black) between January 18 and November 28, 2021. Blue shaded lines mark the periods of change points detected in the CTIS. Black shaded lines mark the periods of change points detected in the CDC benchmark.}
\end{figure}

\pagebreak

\bibliography{main.bib}

\end{document}